\documentclass[]{jfm}
\usepackage{array} 
\usepackage{graphicx}
\usepackage{caption}
\usepackage{subcaption}
\usepackage{tikz} 
\usepackage{algorithm,algorithmic}
\usepackage{newtxtext}
\usepackage{newtxmath}
\usepackage{natbib}
\usepackage{hyperref}
\hypersetup{
    colorlinks = false,
    urlcolor   = blue,
    citecolor  = black,
    hidelinks
}

\newcommand{\RomanNumeralCaps}[1]
\linenumbers

\AtBeginDocument{
    \captionsetup{
        justification=justified,
        width=\linewidth,
        singlelinecheck=false
    }
    \captionsetup[sub]{
        justification=centering
    }
}

\makeatletter
\def\footerflagdefns#1{}
\makeatother

\title{Modelling and synthesizing turbulence with multi-scale coherent vortices}

\author{Zishuo Han\aff{1},
  Weiyu Shen\aff{1,2}
 \and Yue Yang\aff{1,3}
 \corresp{\email{yyg@pku.edu.cn}}
 }

\affiliation{
\aff{1}State Key Laboratory for Turbulence and Complex Systems, School of Mechanics and Engineering Science, Peking University, Beijing 100871, PR China
\aff{2}Max Planck Institute for Solar System Research,  G\"ottingen 37077, Germany
\aff{3}HEDPS-CAPT, Peking University, Beijing 100871, PR China
}

\begin{document}
\maketitle

\begin{abstract}

Turbulence is a complex system exhibiting both universal statistical features and prominent coherent structures. 
We model turbulence using coherent vortices distributed within a multi-scale statistical framework, termed `woven turbulence'.
These entangled vortices are generated based on fractional Brownian bridges, with scale-dependent parameters set by dimensional analysis and geometric similarity.
By integrating statistical and structural modeling, our approach naturally captures both the universal statistical features of turbulence and its coherent vortex structures.
The spatial filling fraction of vortices in woven turbulence, termed `vortex density', is tunable, enabling us to investigate the statistical-structural interaction and uncover two concise physical insights of turbulence. 
First, the invariance of the hierarchical vortex density across scales corresponds to Kolmogorov's $-5/3$ law in the inertial range.
Second, there exists a critical total vortex density at which the intermittency of woven turbulence closely matches that of real turbulence, and this critical density converges to a finite value in the inviscid limit. Deviating from this critical density reveals a negative correlation between intermittency and total vortex density.
In addition, woven turbulence also serves as a fast turbulence synthesis method, requiring only the Taylor-Reynolds number as input and exhibiting an extremely low computational cost proportional to the grid size. 
It generates instantaneous turbulent fields at Taylor-Reynolds numbers of order $10^3$ on $4096^3$ grid points, with computational cost over five orders of magnitude lower than that of direct numerical simulation. 

\end{abstract}

\begin{keywords}
isotropic turbulence, vortex dynamics, turbulence theory
\end{keywords}


\section{Introduction}
Turbulence, prevalent in both natural and engineering contexts, is acknowledged as one of the most complex systems \citep{Davidson2004Turbulence,Benzi2023Lectures,deWit2024,Sreenivasan2025What}.
Understanding its intricate physical mechanisms and achieving efficient simulations remain significant challenges.
Various turbulence modelling methods \citep[e.g.][]{Fung1992Kinematic, Rosales2006A, Lesaffre2025Multiscale, Patruno2018A, Zhou2015Multifractal, Meneveau2000Scale, Kerstein1988A,Arneodo1998Random,Biferale1998Mimicking,Chen2024A,Wu2017Inflow} have been proposed to understand and synthesize turbulence, each with its own strengths and limitations.
  
Statistical modelling and synthesis methods of turbulence incorporate randomness and aim to reproduce major turbulence statistics.
For example, the random Fourier modes method \citep{Kraichnan1970diffusion,Fung1992Kinematic,Patruno2018A} generates Fourier modes constrained by a specified energy spectrum so that it can reproduce low-order statistical features of turbulence.
Although it is widely used to generate initial fields for numerical simulation due to its low computational cost, it cannot reproduce high-order statistics of turbulence. 
Fractal models are based on the concept of energy cascade and characterize turbulence as multi-scale self-similar vortices \citep{Sreenivasan1991Fractals}.
These models, including monofractal (e.g.~the $\beta$-model \citep{Frisch1978A}) and multifractal ones (e.g.~the p-model \citep{Meneveau1987Simple}), can reproduce high-order statistics of turbulence.
Vortices in these fractal models are either conceptual entities, represented as simplified structures such as rectilinear vortex filaments \citep{Chorin1986Turbulence}, or approximately synthesized using mathematical tools like wavelet-based methods \citep{Arneodo1998Random,Malara2016Fast,Zhou2015Multifractal,Zhou2021phdthesis,Robitaille2020Statistical,L2023Stochastic} and tensor trains \citep{Pisoni2025Compression}.
However, vortices in turbulence are governed by complex dynamics, leading to intricate coherent structures that cannot be reproduced by current fractal models.

Coherent vortices are considered the key to understanding the physical mechanisms of turbulence \citep{Pullin1998vortex,Polanco2021Vortex,Cardesa2017The,K1965Report,Yang2023Applications}. 
Specifically, they govern energy transfer \citep{McKeown2020Turbulence,Ostilla2021Cascades} and dissipation \citep{Zinchenko2024Local}, and their interaction with strain \citep{Musci2025Experimental,Zhao2025Evolution} contributes to intermittency \citep{Elsinga2023Intermittency}.
Structural modelling methods study possible flow structures in turbulence starting from the dynamics of the Navier-Stokes (NS) equations \citep{Pullin1998vortex}, offering a physics-based alternative pathway to turbulence modelling.
Several vortex models, such as spherical \citep{Chou1957The,Chou1995review,Synge1943On}, tubular \citep{Townsend1951On}, sheet \citep{Corrsin1962Turbulent}, and spiral vortices \citep{Lundgren1982Strained}, have been proposed to describe the fine-scale structures of turbulence.
Given that vortices in real turbulence exhibit complex, intertwined coherent structures \citep{She1990Intermittent, Jim1993The, Yang2011vsf, xiong2019Identifying}, a critical issue lies in how to intricately organize these fine-scale vortex elements into a turbulent field. 
Early studies on how such vortex structures give rise to turbulence statistics have largely relied on analytical or theoretical approaches. 

Recently, there have been some attempts to model turbulence by organizing specific vortex structures.
While they have successfully explained how statistical features are linked to these vortices, they generally face technical difficulties.
For example, \citet{Schorlepp2025Synthetic} models turbulence using the instanton formalism inspired by quantum and statistical field theory, but this method is currently limited to one-dimensional problems. \citet{Zinchenko2024Local} constructs Burgers vortex tubes to model turbulence, but can only achieve random superpositions of straight vortex tubes. \citet{A2020Vortex,Moriconi2022Circulation,Moriconi2024Vortex} describe turbulence as dilute vortex tubes and model the planar vortices formed by the intersection of vortex tubes with a plane, but are restricted to studying two-dimensional slices.
Therefore, a practical method for constructing three-dimensional customizable vortex structures is needed to provide a testbed for investigating the relationship between vortex structures and statistical features in turbulence.

\citet{xiong2019Construction,xiong2020Effects} developed a method for constructing complex vortex tubes based on explicit vorticity expressions. 
Subsequently, \citet{Shen2024Designing} extended this method and proposed a bottom-up approach to synthesize turbulence by constructing Burgers vortex tubes \citep{Burgers1948A} based on vortex filaments \citep{Barenghi2014Introduction}.
This method simultaneously reproduces the intertwined coherent vortices, key statistics such as Kolmogorov's -5/3 law \citep{Kolmogorov1941The}, and intermittency \citep{She1990Intermittent} observed in real turbulence.
However, in this synthetic turbulence, vortex tube scale varies only within a narrow range, and the overall multi-scale characteristics largely depend on the entanglement properties of vortex filaments \citep{Kivotides2003quantized}.
This introduces several limitations. 
First, representations based solely on vortex filaments fail to capture the intricate multi-scale structure of real turbulence. As a consequence, the generated vortices differ considerably from those in real turbulence.
Second, vortex filaments are generated from computationally demanding dynamical simulations \citep{Hanninen2014vortex} and difficult to control directly.
These limitations motivate a more refined approach that can synthesize turbulence with inherent multi-scale vortex tubes, whose centerlines are both easily adjustable and rapidly generated.

Building upon the work of \citet{Shen2024Designing} and incorporating a fractal framework, this study models turbulence as stochastic fractal vortex tubes with a critical vortex density.
The stochasticity is achieved by generating the vortex  centerline using the fractional Brownian bridge (FBB), a computationally efficient \citep{Dieker2003On, Dieker2004Simulation} stochastic process with adjustable multi-scale features \citep{Friedrich2020Stochastic, Delorme2016Extreme}.
The generated flow field is referred to as `woven turbulence' \citep{Shen2024Designing}, which combines the advantages of statistical and structural modelling methods.
It generates vortex tubes that resemble real turbulence with inherent intermittency that is adjustable via the vortex density, and incorporates multi-scale characteristics and stochasticity from the statistical framework. 
This approach provides a testbed for investigating the statistical-structural relation in turbulence, offering insights into its multi-scale structure and extreme events, and enabling the fast synthesis of realistic turbulent fields. 

The structure of this paper is as follows. 
Section~\ref{sec:methodology} introduces the method for constructing woven turbulence.
Section~\ref{sec:stat} elucidates how the turbulence statistics are generated from the vortices in woven turbulence.
Section~\ref{sec:Synthetic turbulence} assesses the computational efficiency and turbulence characteristics of woven turbulence through a comparison with direct numerical simulation (DNS) results at a range of Taylor-Reynolds numbers $Re_\lambda$. 
Some conclusions are drawn in Section~\ref{sec:conclusion}. 

\section{Construction of woven turbulence}
\label{sec:methodology}
\begin{figure}
  \centering
  \includegraphics[width=1.0\textwidth]{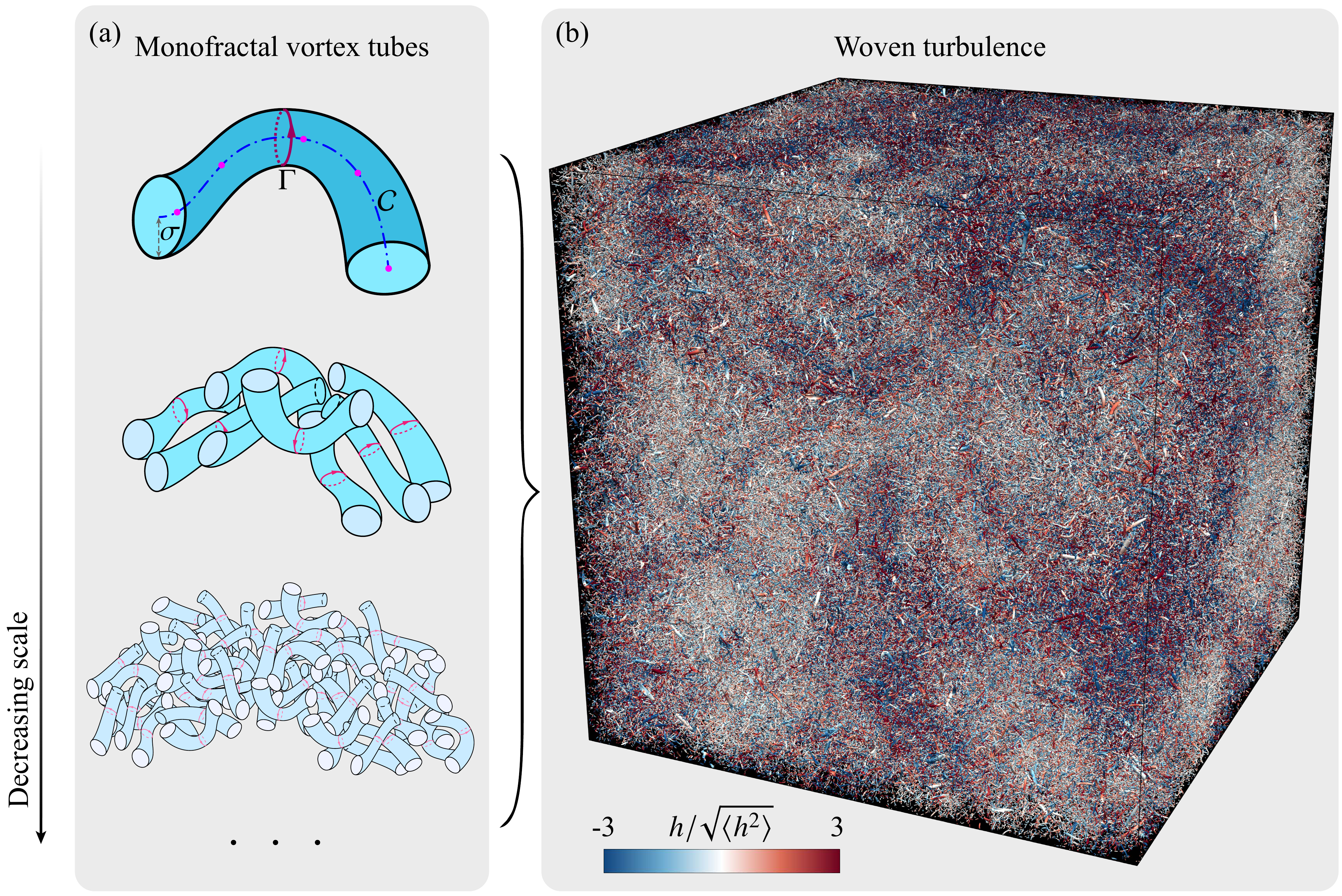}
  \caption{
  Construction of woven turbulence.
  (a) Schematic of multi-scale vortex tubes constituting woven turbulence.
  Each vortex tube is constructed around its curved centerline, and this centerline is generated from spline interpolation of the FBB discrete points.
  At the first scale level, we annotate the three fundamental elements of each vortex tube: the core size $\sigma$, the circulation $\Gamma$, and the centerline $\mathcal{C}$, with the underlying FBB discrete points $\boldsymbol{B}({J})$ indicated as purple points.
  For vortex tubes across different scales, both the centerline length and circulation are scale-dependent but uniform within each scale, whereas centerline shapes vary stochastically.
  Note that the vortex tubes are sketched as segments, but the actual ones in woven turbulence are closed. 
  (b) Visualization of the woven turbulence case WT6 (with $4096^3$ grid points and $Re_{\lambda}=1237$, see table~\ref{tab:set-up of cases}). 
  Flow fields in a $1/2^3$ spatial portion are visualized by isosurfaces of the vorticity magnitude $|\boldsymbol{\omega}|=3\sqrt{\langle\Omega\rangle}$,
  where $ \Omega  = |\boldsymbol{\omega}|^2/2$ denotes the enstrophy and $\langle\cdot\rangle$ represents the volume average over the computational domain ${V}$.
  These isosurfaces are color-coded by the normalized helicity density $h/\sqrt{\langle h^2\rangle}$ with helicity density $h = \boldsymbol{u}\cdot\boldsymbol{\omega}$, where $\boldsymbol{u}$ and $\boldsymbol{\omega}$ denote the velocity and the vorticity, respectively.
  }
\label{fig:multi-scale}
\end{figure}

The construction procedure of woven turbulence is sketched in figure~\ref{fig:multi-scale}. 
The woven homogeneous isotropic turbulence (HIT) is composed of multiple sets of stochastic vortices across $\mathcal{N}$ discrete scale levels within a periodic box ${V}$ of side $\mathscr{L}=2\pi$. 
These vortices are simplified to follow a stochastic monofractal framework as shown in figure~\ref{fig:multi-scale}(a).
The breakup of large-scale vortices into smaller ones resembles a uniform division into multiple equal-sized vortex tubes, with circulation decaying uniformly across scales, while centerline shapes vary stochastically. 
This breakup process mimics the stretching \citep{Johnson2021On,Doan2018Scale} and subsequent vortex breakup in real turbulence.

\subsection{Structural modelling: entangled vortex tubes}
\label{sec:methods_single}

The present construction is based on vortex tubes, which are prevalent coherent structures in turbulence \citep{Pullin1998vortex,Elsinga2017The,Sharma2021Local}, with potential for extension to other structures such as vortex sheets \citep{Shen2024Designing}.
As shown in figure~\ref{fig:multi-scale}(a), each vortex tube is constructed along its curved centerline $\mathcal{C}$, which is generated based on FBB discrete points in 3D space.
The FBB $\boldsymbol{B}({J})$ is a discrete stochastic process with ${J} = 1, 2, ..., \mathscr{N}$ indexing its points and $\mathscr{N}$ denoting the total number of points.
It exhibits Gaussianity and fractal behavior while allowing precise adjustment of long-range correlations via the Hurst exponent $H$ \citep{Friedrich2020Stochastic,Delorme2016Extreme}, and has found wide applications across diverse fields \citep{Wei2000Single, Krapivsky2014Large, Kukla1996NMR, Kassel2024Nonlinear, Molz1997Fractional, Rostek2013A}.
The use of FBB ensures the stochasticity and entanglement of vortex tubes.
 
\begin{figure}
  \centering
  \includegraphics[width=1.0\textwidth]{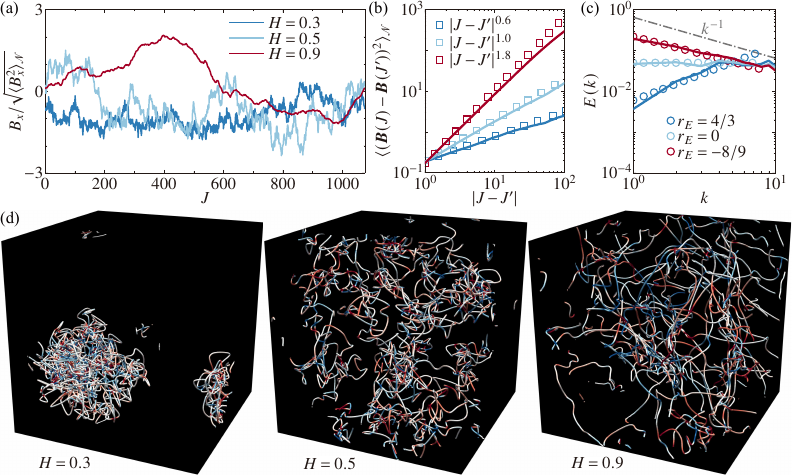}
  \caption{
  Effect of Hurst exponent $H$ on FBB and spatial arrangement of vortex tubes in woven turbulence.
  These cases use the same parameter values $\mathcal{N}=1$, $\sigma_{\mathcal{N}}=0.01$, $\varrho=1.98 \times 10^{-4}$, and $\lambda_{\sigma}=0$, with $H = 0.3$, $0.5$, and $0.9$. 
  (a) Sample path of the FBB $x$-component $B_{x}(J)=\boldsymbol{B}(J)\cdot\boldsymbol{e}_{x}$, normalized by its standard deviation $\sqrt{\langle B_{x}^2 \rangle_{\mathscr{N}}}$.
  (b) Scaling of mean-squared distance $\langle (\boldsymbol{B}({J}) - \boldsymbol{B}({J}^{\prime}))^2 \rangle_{\mathscr{N}}$ in generated FBB (solid lines) and corresponding theoretical scaling law in \eqref{eq:FBB_scaling} (symbols).
  (c) Energy-containing range spectra of woven turbulence cases (solid lines) and corresponding model spectra $E(k) \propto k^{r_{E}}$ (symbols).
  (d) Vorticity magnitude isosurfaces $|\boldsymbol{\omega}|=3\sqrt{\langle\Omega\rangle}$ of woven turbulence cases with $H=0.3$, $0.5$, and $0.9$ from left to right.
  These isosurfaces are color-coded by the normalized helicity density $h/\sqrt{\langle h^2\rangle}$ with the colorbar same as that in figure~\ref{fig:multi-scale}(b).
  }
  \label{fig:FBB}
\end{figure}

For each vortex tube with core size $\sigma_{i}$ characterizing its scale, we first generate the corresponding FBB (shown as purple points in figure~\ref{fig:multi-scale}(a)) by wavelet-based synthesis \citep{Abry1996the,Bardet2003theory}.
Figure~\ref{fig:FBB}(a) shows normalized sample paths of 1D FBBs with varying $H \in (0,1)$.
The Hurst exponent $H$ affects the long-range dependence of FBB, e.g.~a smaller $H$ means stronger fluctuations of FBB as in figure~\ref{fig:FBB}(a). 
For FBB with indices ${J}$ and ${J}^{\prime}$, the mean-squared distance satisfies the scaling \citep{Friedrich2020Stochastic} 
\begin{equation}
    \label{eq:FBB_scaling}
    \langle (\boldsymbol{B}({J}) - \boldsymbol{B}({J}^{\prime}))^2 \rangle_{\mathscr{N}} = \delta_B^2 |{J}-{J}^{\prime}|^{2H}
\end{equation}
theoretically, where $\langle\cdot\rangle_{\mathscr{N}}$ denotes the average over $\mathscr{N}$ points and $\delta_B$ the FBB scale factor.
Figure~\ref{fig:FBB}(b) confirms that the generated FBB satisfies the scaling in \eqref{eq:FBB_scaling}.
To link the FBB scale to the vortex tube, we set $\delta_B = 40 \sigma_{i}$ for each tube.
Note that the FBB is end-to-end connected in this work to ensure the closure of the vortex tube. 
The effect of FBB on the woven turbulence will be discussed in Section~\ref{sec:Energy-containing and dissipation range}.
Next, these discrete points of FBB are interpolated by the fifth-order splines and mapped into the periodic box ${V}$ by taking the modulo with respect to the box side length $\mathscr{L}$. 
Then we obtain the vortex centerline $\mathcal{C}$ (shown as the blue dashed-dot line in figure~\ref{fig:multi-scale}(a)) with position $\boldsymbol{c}(s)$ along arc-length $s$.
The length of centerline 
\begin{equation}
\label{eq:centerline_length}
    L = \int_{\mathcal{C}} \mathrm{d}s \approx c_{H} \delta_B \mathscr{N}
\end{equation}
is proportional to $\mathscr{N}$ and $\delta_B$, where the value of the coefficient $c_{H}$ is determined by $H$. 

Subsequently, the vortex tube is generated based on its centerline $\mathcal{C}$. 
The curved cylindrical coordinate system $(s, \rho, \theta)$ is introduced with the local cylindrical frame $(\boldsymbol{e}_s,\boldsymbol{e}_\rho,\boldsymbol{e}_\theta)$ in a tubular region surrounding centerline $\mathcal{C}$ with radius $\mathcal{R}$. 
The system satisfies $\boldsymbol{x}(s, \rho, \theta)=\boldsymbol{c}(s)+\rho \boldsymbol{e}_\rho(\theta)$.
At the scale level with index ${i} \in \{1, 2, \ldots, \mathcal{N}\}$, the vorticity of the ${j}$-th vortex tube is specified by \citep{Shen2024Designing,Shen2023Role}
\begin{equation}
\label{eq:w_single}
    \boldsymbol{\omega}_{ij}(\boldsymbol{x}(s, \rho, \theta))=\Gamma_{i} G_{i}(s, \rho)\left[{\boldsymbol{e}_s}
    +{\frac{\rho}{R_{i}(s)(1-\kappa(s) \rho \cos \theta)} \frac{\mathrm{d} R_{i}(s)}{\mathrm{d} s} \boldsymbol{e}_\rho}
    \right]
\end{equation}
with the circulation $\Gamma_{i}$, the curvature $\kappa$, the Gaussian kernel function
\begin{equation}
\label{eq:kernel}
    G_{i}(s, \rho)
    = \begin{cases}
    \frac{1}{2 \pi R_{i}(s)^2} \exp \left[\frac{-\rho^2}{2 R_{i}(s)^2}\right], \quad \rho \in\left[0, \mathcal{R}\right), 
    \\ 0, \quad \rho \in\left[\mathcal{R},+\infty\right)\end{cases}
\end{equation}
as in the vortex model of \citet{Burgers1948A}, 
and the local vortex core size 
\begin{equation}
    \label{eq:core size}
    R_{i}(s) = \sigma_{i}\left(1+\lambda_{\sigma}\left(1+\sin\left(2\pi M_{i}{{s}}/{L_{i}}\right)\right)\right).
\end{equation}
Here, the core size $\sigma_{i}$ represents the characteristic scale of the vortex tube at the $i$-th scale level, $\lambda_{\sigma}\geq 0$ denotes the core size variation magnitude and $M_{i}$ the variation frequency.
Note that all vortex tubes at the $i$-th scale level share the same parameters $(\Gamma_i, \sigma_i, M_i)$ and differ only in their centerlines.

\subsection{Statistical modelling: fractal framework}
\label{sec:Fractal framework}

The vorticity of woven turbulence
\begin{equation}
\label{eq:w_sum}
\boldsymbol{\omega}=\sum_{{i}=1}^{\mathcal{N}}\boldsymbol{\omega}_{i}=\sum_{{i}=1}^{\mathcal{N}}\sum_{{j}=1}^{n_{i}}\boldsymbol{\omega}_{i{j}}
\end{equation}
is given by the sum of vorticity over all vortex tubes, where $n_{i}$ denotes the population of vortex tubes at the $i$-th scale level, and $\boldsymbol{\omega}_{i}=\sum_{{j}=1}^{n_{i}}\boldsymbol{\omega}_{i{j}}$ denotes the total vorticity of vortex tubes at the $i$-th scale level. 
The core size $\sigma_{i}$, centerline length $L_{i}$, population $n_{i}$, and circulation $\Gamma_{i}$ of vortex tubes at different scales are set to satisfy self-similar relations
\begin{equation}
\label{eq:self-similar}
    \begin{cases}
    \sigma_{i+1}/\sigma_{i} = r_{\sigma}, \\
    L_{i+1}/L_{i} = r_{L}, \\
    n_{i+1}/n_{i} = r_{n}, \\
    \Gamma_{i+1}/\Gamma_{i} = r_{\Gamma}
    \end{cases}
\end{equation}
for ${i} \in \{1, 2, \ldots, \mathcal{N}-1\}$, where $r_{\sigma}$ denotes the core size ratio, $r_{L}$ the centerline length ratio, $r_{n}$ the population ratio, and $r_{\Gamma}$ the circulation ratio.
In addition, the core size variation frequency $M_{i} = \left\lfloor{2 L_{i}}/{\mathscr{L}}\right\rceil/r_{\sigma}^{i-1}$ in \eqref{eq:core size} is set to ensure the similarity among vortex structures at different scales, where $\lfloor\cdot\rceil$ denotes the rounding operation.
Generally, we set $r_{\sigma}=1/2$ and $n_{1}=1$ as in typical fractal models \citep{Frisch1978A,Meneveau1987Simple,Zhou2021phdthesis}.

In terms of circulation, we assume $\Gamma_{i}$ of a vortex tube with core scale $\sigma_{i}$ depends only on mean dissipation rate $\langle\epsilon\rangle$ and $\sigma_{i}$.
Dimensional analysis suggests $\Gamma_{i}  \propto \sigma_{i}^{4/3}$, so we set $r_{\Gamma} = r_{\sigma}^{4/3} = (1/2)^{4/3}$.
This ratio is consistent with the DNS results \citep{Iyer2019Circulation}.
The circulation ratio $r_{\Gamma}=(1/2)^{4/3}<1$ corresponds to the dissipation of vortex tubes during their evolution toward small scales in viscous flows, contrasting with circulation conservation in inviscid flows.
Small-scale vortex tubes have stronger peak vorticity than large-scale ones since $r_{\Gamma}/r_{\sigma}^2>1$, corresponding to the local vorticity accumulation at small scales.
Note that $r_{\Gamma}$ here characterizes only the circulation statistics in the inertial range, while the circulation statistics in the dissipation range also depends on the vortex structure.

In terms of geometry, the length of the vortex tubes scales proportionally with their core size \citep{Ghira2022Characteristics}, yielding $r_{L}=r_{\sigma}=1/2$.
To quantify how much physical space the vortices occupy, we introduce the total vortex density
\begin{equation}
\label{eq:vortex density}
    \varrho 
    = \sum_{i=1}^{\mathcal{N}} \varrho_{i} 
    = \sum_{i=1}^{\mathcal{N}} \frac{n_{i}L_{i}\sigma_{i}^2}{\mathscr{L}^3}
\end{equation}
with the hierarchical vortex density
\begin{equation}
\label{eq:hierarchical vortex density}
    \varrho_{i}
    = \frac{n_{i}L_{i}\sigma_{i}^2}{\mathscr{L}^3}
    = \frac{n_{1}L_{1}\sigma_{1}^2}{\mathscr{L}^3} (r_{n}r_{L}r_{\sigma}^2)^{i-1}
\end{equation}
characterizing the abundance of vortices at the $i$-th scale level.

The specific settings of woven turbulence cases in this study are provided in table~\ref{tab:set-up of cases}. 
The construction of woven turbulence is governed by the range of scales from $\sigma_{1}$ to $\sigma_{\mathcal{N}}$ in \eqref{eq:core size}, alongside two parameters constrained by universal statistical features: the population ratio $r_{n}$ in \eqref{eq:self-similar} and the total vortex density $\varrho$ in \eqref{eq:vortex density}. 
The values of $\sigma_{1}$, $\sigma_{\mathcal{N}}$, $r_{n}$ and $\varrho$ will be determined, and their effects on turbulence statistics and structures will be discussed in the next section. 
Note that the construction algorithm is detailed in Appendix~\ref{app:Numerical}, and the corresponding code is available at https://github.com/YYgroup/FastWeavTurb.

\begin{table}
  \begin{center}
  \setlength{\tabcolsep}{10pt}
\def~{\hphantom{0}}
  \begin{tabular}{lcccccc}
       Case  & $\mathcal{N}$ & $\sigma_{\mathcal{N}}$  & $Re_{\lambda}$ & $r_{n}$ & $\varrho$ & $N$ \\[3pt]
        WT1  & 2 & $3.68\times 10^{-2}$ & 101 & 8 & $3.40\times 10^{-2}$ & 128 \\
        WT2  & 3 & $1.51\times 10^{-2}$ & 159 & 8 & $2.40\times 10^{-2}$ & 256 \\
        WT3  & 4 & $7.80\times 10^{-3}$ & 268 & 8 & $1.70\times 10^{-2}$ & 512 \\
        WT4  & 5 & $3.57\times 10^{-3}$ & 419 & 8 & $1.28\times 10^{-2}$ & 1024 \\
        WT5  & 6 & $1.70\times 10^{-3}$ & 722 & 8 & $1.21\times 10^{-2}$ & 2048 \\
        WT6  & 7 & $8.47\times 10^{-4}$ & 1237 & 8 & $1.20\times 10^{-2}$ & 4096 \\
        WT2-I   & 3 & $1.51\times 10^{-2}$ & 149 & 8 & $2.40\times 10^{-3}$ & 512 \\
        WT2-C   & 3 & $1.51\times 10^{-2}$ & 159 & 8 & $2.40\times 10^{-2}$ & 512 \\
        WT2-E   & 3 & $1.51\times 10^{-2}$ & 152 & 8 & $2.40\times 10^{-1}$ & 512 \\
        WT4-S  & 5 & $3.57\times 10^{-3}$ & - & 4 & $4.96\times 10^{-3}$ & 1024 \\
        WT4-D  & 5 & $3.57\times 10^{-3}$ & - & 16 & $7.94\times 10^{-2}$ & 1024 \\
    \end{tabular}
  \caption{
  Setup of woven turbulence cases. 
  The cases from WT1 to WT6 are designed to investigate woven turbulence across a wide range of Taylor-Reynolds numbers $Re_{\lambda}$ from $O(10^2)$ to $O(10^3)$. 
  For these cases, the total vortex density $\varrho$ defined in \eqref{eq:vortex density} is set to the critical value $\varrho_c$. 
  The WT2-based variants (WT2-I, WT2-C, and WT2-E) are used to study the effects of total vortex density $\varrho$, with $\varrho$ spanning three orders of magnitude while keeping other parameters constant.
  The WT4-based variants (WT4-S and WT4-D) are designed to investigate the effect of the population ratio $r_{n}$ in \eqref{eq:self-similar}. 
  The hierarchical vortex density at the largest scale, $\varrho_1$ defined in \eqref{eq:hierarchical vortex density}, is held constant in these cases. 
  All cases have the same root-mean-square velocity $u^{\prime}=1$. 
  }
  \label{tab:set-up of cases}
  \end{center}
\end{table}

\section{Structure-based modelling of woven turbulence statistics}
\label{sec:stat}

\subsection{Energy spectrum}
\label{sec:energy spectrum}

\subsubsection{Inertial range}
\label{sec:Inertial range}
In general, the energy spectrum of turbulence exhibits an energy-containing range, an inertial range, and a dissipation range.
These three regions correspond to distinct physical mechanisms at different scales.
The inertial range is central to turbulence theory, where scale-invariant dynamics and energy cascades give rise to universal statistical laws, exemplified by the $-5/3$ power law of \cite{Kolmogorov1941The}.
This universality is often explained by the self-similar organization of multi-scale vortices.

\begin{figure}
    \centering
       \includegraphics[width=1.0\textwidth]{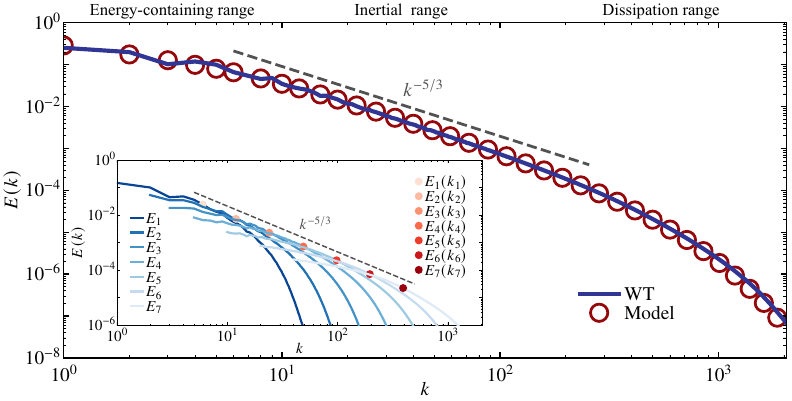}
    \caption{
    Energy spectrum $E(k)$ of case WT6 (see table~\ref{tab:set-up of cases}) with $Re_{\lambda}=1237$ and the corresponding model spectrum in \eqref{eq:model_spectrum}. 
    The inset plots the hierarchical energy spectra $E_{i}(k)$ in \eqref{eq:hierarchical energy spectrum} of case WT6 and the corresponding characteristic energy $E_{i}(k_{i})$.
    Each hierarchical energy spectrum is shown around its characteristic wavenumber $k_i$, within the range $[k_i/10, 10k_i]$. 
    }
    \label{fig:Energy spectra}
\end{figure}

Accordingly, we generate the multi-scale self-similar vortices through \eqref{eq:w_sum}, \eqref{eq:w_single} and \eqref{eq:self-similar}, which give rise to the inertial range in woven turbulence. 
To investigate the relation between the vortex parameters and the scaling exponent of the inertial-range spectrum, 
we define the hierarchical energy spectrum
\begin{equation}
    \label{eq:hierarchical energy spectrum}
    E_{i}(k) = \oint_{S(k)}\frac{1}{2}\hat{\boldsymbol{u}}_{i}(\boldsymbol{k})\overline{\hat{\boldsymbol{u}}}_{i}(\boldsymbol{k})\mathrm{d}S(k).
\end{equation}
Here $\hat{\boldsymbol{u}}_{i}(\boldsymbol{k})={i \boldsymbol{k} \times (\sum_{{j}=1}^{n_{i}}\widehat{\boldsymbol{\omega}}_{i}}(\boldsymbol{k}))/{k^2}$ denotes the velocity at the $i$-th scale level in Fourier space, $S(k)$ the sphere with a radius of $k$ in Fourier space, $\boldsymbol{k}$ the wavevector, $k=|\boldsymbol{k}|$ the wavenumber, and the overline represents the complex conjugate.  
Figure~\ref{fig:Energy spectra} plots the total and hierarchical energy spectra for case WT6. 
The hierarchical spectra at different scales exhibit a similar shape, consistent with the self-similar nature of the fractal vortices, and support the presence of an inertial range in woven turbulence. 
Since $E_i(k)$ spans a range of wavenumbers, for simplicity, we extract a representative value for each scale by introducing a characteristic energy $E_{i}(k_{i})$, where ${k}_{i} = {1}/(3\sigma_{i})$ denotes the corresponding characteristic wavenumber.
In Figure~\ref{fig:Energy spectra}, $E_i(k_i)$ and $E(k)$ both follow a $k^{-5/3}$ scaling in the inertial range. 
This agreement suggests the \textit{hierarchical energy hypothesis}
\begin{equation}
\label{eq:hyp_1}
\begin{split}
\frac{E(k_{{i}+1})}{E(k_{i})}
= \frac{E_{{i}+1}(k_{{i}+1})}{E_{i}(k_{i})},
\end{split}
\end{equation}
i.e.~the scaling of $E(k)$ in woven turbulence can be captured through the ratio of hierarchical energies at adjacent scale levels. 

Next, we investigate how the hierarchical energy spectrum is determined by the parameters of vortex tubes. 
Since 
\begin{equation}
\label{eq:spectrum via vorticity}
    E_{i}({k})
    = \oint_{S(k)}\frac{1}{2{k}^2}\hat{\boldsymbol{\omega}}_{i}(\boldsymbol{k})\overline{\hat{\boldsymbol{\omega}}}_{i}(\boldsymbol{k})\mathrm{d}S(k)
\end{equation} 
can be expressed in terms of the vorticity $\hat{\boldsymbol{\omega}}_i(\boldsymbol{k})$ in Fourier space \citep{Davidson2004Turbulence}, we analyze how $\hat{\boldsymbol{\omega}}_i(\boldsymbol{k})$ of vortex tubes at each scale can be expressed in terms of vortex tube parameters.
We first define the
normalized arc-length $s^{*} = s/\sigma_{i}$, radius $\rho^{*} = \rho/\sigma_{i}$, position vector $\boldsymbol{x}^{*}(s^{*},\rho^{*},\theta) = \boldsymbol{x}(s,\rho,\theta)/\sigma_{i}$ and wavevector $\boldsymbol{k}^{*} = \boldsymbol{k}\sigma_{i}$. 
They allow us to express
\begin{equation}
\label{eq:vorticity_Fourier}
\begin{split}
\hat{\boldsymbol{\omega}}_{i}(\boldsymbol{k})
& 
=\frac{1}{2\pi}\sum_{{j}=1}^{n_{i}}\int_{{V}_{ij}} \boldsymbol{\omega}_{ij}(\boldsymbol{x}) e^{-\mathrm{i}\boldsymbol{k}\cdot\boldsymbol{x}}\mathrm{d}\boldsymbol{x}
\\
& 
=\frac{1}{2\pi}\sum_{{j}=1}^{n_{i}}\int_{{V}_{ij}} \boldsymbol{\omega}_{ij}(\boldsymbol{x}(s,\rho,\theta)) e^{-\mathrm{i}\boldsymbol{k}\cdot\boldsymbol{x}(s,\rho,\theta)}\rho\mathrm{d}\theta\mathrm{d}\rho\mathrm{d}s
\\
& 
=\frac{1}{2\pi}\sum_{{j}=1}^{n_{i}}\int_{{V}_{ij}} \frac{\Gamma_{i}\sigma_{i}}{2\pi}\frac{{\sigma}^2_{i}}{{R}^2_{i}}\exp\left(-\frac{\rho^{*2}}{2}\frac{{\sigma}^2_{i}}{{R}^2_{i}}\right)\boldsymbol{e}_{ij} e^{-\mathrm{i}\boldsymbol{k}^{*}\cdot\boldsymbol{x}^{*}(s^{*},\rho^{*},\theta)}\rho^{*}\mathrm{d}\theta\mathrm{d}\rho^{*}\mathrm{d}s^{*}
\\
& 
=\frac{\Gamma_{i}\sigma_{i}}{2\pi} \sum_{{j}=1}^{n_{i}}\mathcal{W}_{ij}(\boldsymbol{k}^{*}),
\end{split}
\end{equation}
where ${V}_{ij}$ denotes the space occupied by the $j$-th tube at the $i$-th scale level, $\boldsymbol{e}_{ij} = \boldsymbol{\omega}_{ij}/|\boldsymbol{\omega}_{ij}|$ the direction vector, and
\begin{equation}
\label{eq:characteristic_vorticity_Fourier}
\mathcal{W}_{ij}(\boldsymbol{k}^{*}) = \frac{1}{2\pi}\int_{{V}_{ij}} \frac{{\sigma}^2_{i}}{{R}^2_{i}}\exp\left(-\frac{\rho^{*2}}{2}\frac{{\sigma}^2_{i}}{{R}^2_{i}}\right)\boldsymbol{e}_{ij} e^{-\mathrm{i}\boldsymbol{k}^{*}\cdot\boldsymbol{x}^{*}(s^{*},\rho^{*},\theta)}\rho^{*}\mathrm{d}\theta\mathrm{d}\rho^{*}\mathrm{d}s^{*}
\end{equation}
the normalized vorticity of the $j$-th tube at the $i$-th scale level in Fourier space. 

Substituting \eqref{eq:vorticity_Fourier} into \eqref{eq:spectrum via vorticity} yields
\begin{equation}
\label{eq:Ei_ki_1}
    E_{i}({k}_{i})
    = \oint_{S(k_{i})}\frac{1}{2{k}_{i}^2}\hat{\boldsymbol{\omega}}_{i}(\boldsymbol{k}_{i})\overline{\hat{\boldsymbol{\omega}}}_{i}(\boldsymbol{k}_{i})\mathrm{d}S(k_{i})
    = \left(\frac{\Gamma_{i}\sigma_{i}}{2\pi}\right)^2 \mathcal{E}_{i}(k_{i}^{*})
\end{equation}
with the normalized characteristic energy
\begin{equation}
\label{eq:norm_char_energy}
\mathcal{E}_{i}(k_{i}^{*}) = \oint_{S(k_{i}^{*})} \frac{1}{2(k_{i}^{*})^2} 
\left( \sum_{j=1}^{n_i} \mathcal{W}_{ij}(\boldsymbol{k}_{i}^{*}) \right)
\left( \sum_{j'=1}^{n_i} \overline{\mathcal{W}}_{ij'}(\boldsymbol{k}_{i}^{*}) \right)
\mathrm{d}S(k_{i}^{*})
\end{equation}
of vortex tubes at the $i$-th scale level.
In \eqref{eq:norm_char_energy} with \eqref{eq:characteristic_vorticity_Fourier}, $\mathcal{E}_{i}(k_{i}^{*})$ depends on the population $n_i$, the normalized centerline length $L_{i}^{*} = L_{i}/\sigma_{i}$ and the geometric shape of vortex tubes. 
Since the centerline shape determined by FBB and the vorticity profile given by \eqref{eq:kernel} are both self-similar and independent of $n_i$ and $L_{i}^{*}$, their contribution can be factored into a universal coefficient $\mathbb{W}$.
This separation of scale-invariant and scale-dependent factors motivates the \textit{vortex-length-dependent energy hypothesis}: 
the normalized characteristic energy  
\begin{equation}
\label{eq:hyp_2}
\mathcal{E}_{i}(k_{i}^{*})= {\mathbb{W}}L_{i}^{*}n_{i}. 
\end{equation}
of vortex tubes at the $i$-th scale level is proportional to $L_{i}^{*}$ and $n_{i}$.  
Based on \eqref{eq:Ei_ki_1} and \eqref{eq:hyp_2}, the relation between the characteristic energy and parameters of
vortex tubes is
\begin{equation}
\label{eq:Ei_ki}
    E_{i}({k}_{i})
    = \left(\frac{\Gamma_{i}\sigma_{i}}{2\pi}\right)^2 \mathcal{E}_{i}(k_{i}^{*})
    = \left(\frac{\Gamma_{i}\sigma_{i}}{2\pi}\right)^2{\mathbb{W}}L_{i}^{*}n_{i}
    = \frac{\mathbb{W}}{(2\pi)^2}\Gamma_{i}^2L_{i}n_{i}\sigma_{i}.
\end{equation}

Finally, substituting \eqref{eq:Ei_ki} into \eqref{eq:hyp_1} with \eqref{eq:self-similar} yields the scaling exponent
\begin{equation}
\label{eq:r_I}
\begin{split}
    r_{I} 
    &
    = \frac{\mathrm{ln}(E_{{i}+1}({k}_{{i}+1})/E_{i}({k}_{i}))}{\mathrm{ln}({k}_{{i}+1}/{k}_{i})}
    \\
    &
    = -\frac{2\mathrm{ln}(\Gamma_{{i}+1}/\Gamma_{i})+\mathrm{ln}(\sigma_{{i}+1}/\sigma_{i})+\mathrm{ln}(L_{{i}+1}/L_{i})+\mathrm{ln}(n_{{i}+1}/n_{i})}{\mathrm{ln}({\sigma}_{{i}+1}/{\sigma}_{i})}
    \\
    &
    =-1-\frac{\mathrm{ln}(r_{n}r_{L})+2\mathrm{ln}(r_{\Gamma})}{\mathrm{ln}(r_{\sigma})}
\end{split}
\end{equation}
in the inertial-range spectrum $E(k) \propto k^{r_{I}}$.
In Appendix~\ref{app:inertial range}, numerical experiments show that \eqref{eq:r_I} is valid for $r_{I} < -1$. 
As established in Section~\ref{sec:Fractal framework}, the core size ratio $r_{\sigma}=1/2$, the centerline length ratio $r_{L}=1/2$, and the circulation ratio $r_{\Gamma}=(1/2)^{4/3}$ are set accordingly.
Substituting these values into \eqref{eq:r_I}, we obtain $r_{I} = -14/3+{\ln r_{n}}/{\ln 2}$.
Therefore, Kolmogorov's -5/3 law \citep{Kolmogorov1941The} $r_{I} = -5/3$ suggests $r_{n}=8$ in woven turbulence.

\begin{figure}
    \centering
       \includegraphics[width=1.0\textwidth]{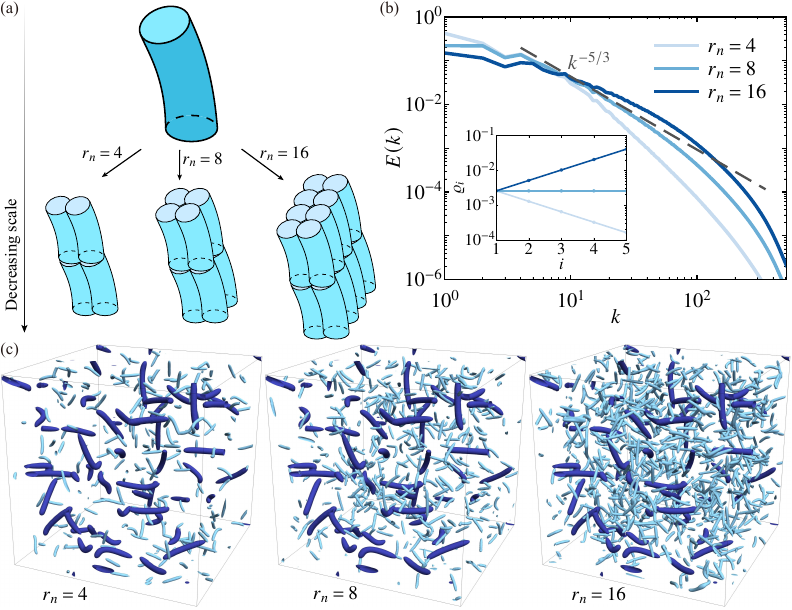}
    \caption{
    Effect of the population ratio $r_{n}$ on woven turbulence.
    Cases WT4-S, WT4, and WT4-D (see table~\ref{tab:set-up of cases}) vary $r_{n}$ with values of 4, 8, and 16, respectively, while keeping the vortices at the first scale level fixed.
    (a) Schematic of monofractal vortex tubes for different $r_{n}$. 
    For $r_{n} = 8$, the ``total volume'' of vortex tubes at each scale is scale-invariant, while it decreases and increases with decreasing scale for $r_{n} < 8$ and $r_{n} > 8$, respectively. 
    (b) Energy spectra with varying $r_{n}$. 
    The variation of the hierarchical vortex density $\varrho_i$ with scale level $i$ is shown in the inset for various $r_{n}$.
    (c) Vorticity magnitude isosurfaces $|\boldsymbol{\omega}|=\Gamma_{i}/(8\pi\sigma_{i}^2)$ of the vortex tubes at the first (blue) and second (cyan) scale levels for various $r_{n}$. 
    }
    \label{fig:schematic_Df}
\end{figure}

To explore the implications of $r_{n} = 8 = (1/r_{\sigma})^3$, we construct woven turbulence with $r_{n} = 4$, 8 and 16. 
Figure~\ref{fig:schematic_Df} illustrates the influence of $r_{n}$ on the spatial distribution of vortices and the energy spectrum. 
Figure~\ref{fig:schematic_Df}(a) sketches how vortex tubes vary across scales for different $r_{n}$, showing that the relative ``total volume'' of vortex tubes at each scale is governed by $r_{n}$. 
For $r_{n} = 8$, the variation of vortices across scales resembles the uniform subdivision of a material volume in 3D space, thereby preserving the total volume across scales. 
This conservation can be expressed by the scale invariance of the hierarchical vortex density
\begin{equation}
\label{eq:scale-invariant vortex density}
    \varrho_{i}
    = \varrho_{1}(r_{n}r_{L}r_{\sigma}^2)^{i-1}
    = \varrho_{1}
    = \frac{\varrho}{\mathcal{N}}
\end{equation}
based on \eqref{eq:hierarchical vortex density}. It implies a self-similar spatial distribution of vortices across scales. 
Figure~\ref{fig:schematic_Df}(c) displays vorticity magnitude isosurfaces of the vortex tubes at the first and second scale levels for various $r_{n}$, illustrating the variation of $\varrho_{i}$ across scales in woven turbulence.
Thus, our model suggests that Kolmogorov's -5/3 law corresponds to the scale invariance of $\varrho_{i}$. 
In addition, as shown in figure~\ref{fig:schematic_Df}(b), when $\varrho_i$ decreases or increases with decreasing scale, the scaling exponent $r_I$ in the inertial range becomes smaller or larger than $-5/3$, respectively.

\subsubsection{Energy-containing and dissipation ranges}
\label{sec:Energy-containing and dissipation range}
The vortex centerlines are generated by FBB, and the long-range dependence of FBB is determined by the Hurst exponent $H$ in \eqref{eq:FBB_scaling}. To investigate the influence of $H$ on the spatial arrangement of vortex tubes, we construct woven turbulence cases with $H = 0.3$, $0.5$ and $0.9$, along with the same $\mathcal{N}=1$, $\sigma_{\mathcal{N}}=0.01$, $\varrho=1.98 \times 10^{-4}$, and $\lambda_{\sigma}=0$. 
The energy spectra and vorticity isosurfaces of these cases are shown in figures~\ref{fig:FBB}(c) and \ref{fig:FBB}(d), respectively. 
As $H$ increases, the centerlines of vortex tubes become smoother and more unfolded in figure~\ref{fig:FBB}(d), and the amplitude of large-scale energy spectrum is increased in figure~\ref{fig:FBB}(c).  

As derived in Appendix~\ref{app:energy-containing range}, the spectrum in the energy-containing range satisfies
$
  E(k) \propto k^{r_{E}}
$
with the scaling exponent
\begin{equation}
\label{eq:r_E H}
    r_{E} = \frac{1}{H}-2
\end{equation}
negatively correlated with $H$, illustrated in figure~\ref{fig:FBB}(c).
Thus, the role of $H$ in controlling the spatial arrangement of vortex centerlines and thereby shaping the energy spectrum in this range is analogous to that of external forcing in real turbulence.
Here we set $H = 5/6$, corresponding to $r_{E} = -4/5$, which is consistent with the scaling in the woven turbulence in \cite{Shen2024Designing}.

In the dissipation range of turbulence,
the energy no longer follows a power-law scaling  \citep{Buaria2020Dissipation}. 
In woven turbulence, the building blocks are curved vortex tubes, whose internal structure, given in \eqref{eq:kernel} and \eqref{eq:core size}, is identical to that of the Burgers vortex, an exact solution of the NS equations. 
This naturally incorporates viscous effects at small scales. 
The uniform Burgers vortex tube, corresponding to $\lambda_{\sigma} = 0$ in \eqref{eq:core size}, yields a Gaussian decay energy spectrum $E(k) \sim e^{-k^2}$ \citep{Pullin1998vortex}.
However, the energy spectrum of real turbulence typically follows exponential decay in the dissipation range \citep{Pope2000Turbulent,Khurshid2018Energy}.

\begin{figure}
  \centering
  \includegraphics[width=1.0\textwidth]{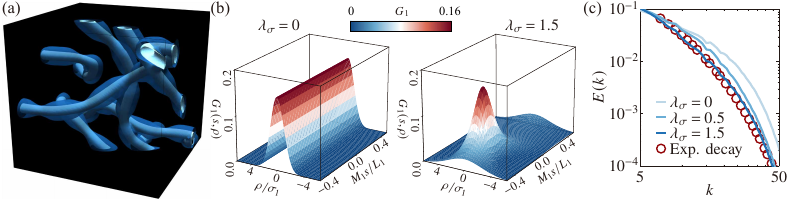}
  \caption{
  Effect of $\lambda_{\sigma}$ on vortex structures and dissipation range energy spectrum in woven turbulence with varying  $\lambda_{\sigma} = 0$, $0.5$, and $1.5$, along with the same parameter values $\mathcal{N}=1$, $\sigma_{\mathcal{N}}=0.04$, $\varrho=1.59 \times 10^{-3}$, and $H=5/6$. 
  (a) Vorticity magnitude isosurfaces $|\boldsymbol{\omega}|=10$ for $\lambda_{\sigma}=0$ (light blue) and $\lambda_{\sigma}=3/2$ (translucent dark blue).
  A $1/2^3$ subdomain of the flow field is shown to highlight the structure in the dissipation range.
  (b) Kernel function $G_{1}(s,\rho)$ given by \eqref{eq:kernel} for $\lambda_{\sigma}=0$ (left) and $\lambda_{\sigma}=3/2$ (right).
  (c) Dissipation range spectra of cases (solid lines) with varying $\lambda_{\sigma}$ and exponential decay model (symbols).
  }
  \label{fig:lambda_disp}
\end{figure}

To address this discrepancy, we treat the core size variation magnitude $\lambda_{\sigma}$ in \eqref{eq:core size} as a tunable parameter, which enables modulation of the fine-scale vortex structure. 
We generate woven turbulence cases with $\lambda_{\sigma} = 0$, $0.5$, and~$1.5$, along with the same $\mathcal{N}=1$, $\sigma_{\mathcal{N}}=0.04$, $\varrho=1.59 \times 10^{-3}$, and $H=5/6$. 
Figures~\ref{fig:lambda_disp}(a), (b), and (c) show the vorticity isosurfaces, the kernel function $G_{1}(s, \rho)$ given in \eqref{eq:kernel}, and the energy spectrum in the small-scale range, respectively, for these cases.
As $\lambda_{\sigma}$ increases from zero, the variation magnitude of core size increases in figure~\ref{fig:lambda_disp}(a), the vorticity governed by $G_{1}(s, \rho)$ in \eqref{eq:w_single} becomes more concentrated in figure~\ref{fig:lambda_disp}(b), and the decay of $E(k)$ at small scales becomes more gradual in figure~\ref{fig:lambda_disp}(c). 
Thus the variation of the vortex core along the centerline introduces non-uniformity in the local dissipation scale, enabling gradual energy decay towards the smallest scale.
Based on these observations, we set $\lambda_{\sigma}=3/2$ to make the dissipation range spectrum approach exponential decay. 

\subsubsection{Reynolds number and model spectrum}
\label{sec:Reynolds number and model spectrum}
Similar to real turbulence, the wavenumber ranges of the three-regime energy spectrum are determined by characteristic length scales of vortices in woven turbulence. 
By matching the energy spectra of woven turbulence and real turbulence across a range of $Re_{\lambda}$ from $O(10^2)$ to $O(10^3)$, we establish scale relations for the integral and Kolmogorov scales
\begin{equation}
\label{eq:scales}
     \mathcal{L}=c_{\mathcal{L}}\sigma_{1} \quad \textrm{and} \quad
    \eta = c_{\eta} \sigma_{\mathcal{N}},
\end{equation}
respectively. Here, $\sigma_{1}$ and $\sigma_{\mathcal{N}}$ are the core sizes of the largest and smallest vortices, respectively, and the coefficients $c_{\mathcal{L}} = 20$ and $c_{\eta} = 0.59$ are calibrated from this spectral agreement.
Considering the Kolmogorov length scale $\eta = ({\nu^3}/{\langle \varepsilon \rangle})^{{1}/{4}}$ and the mean dissipation rate $\langle \varepsilon \rangle = 2 \nu  \langle\Omega\rangle $ \citep{Frisch1995Turbulence}, the effective kinematic viscosity in woven turbulence is estimated by $\nu = \eta^2\sqrt{ 2 \langle\Omega\rangle }$. 
With \eqref{eq:scales}, the Taylor-Reynolds number reads
\begin{equation}
\label{eq:Re_lambda}
    Re_{\lambda} = \frac{u^{\prime}\lambda_{T}}{\nu} = \frac{\sqrt{15}(u^{\prime})^2}{2\eta^2 \langle\Omega\rangle } = \frac{\sqrt{15}(u^{\prime})^2}{2(c_{\eta}\sigma_{\mathcal{N}})^2 \langle\Omega\rangle },
\end{equation}
where $\lambda_{T} = u^{\prime}\sqrt{{15}/(2\langle\Omega\rangle)}$ denotes the Taylor micro-scale, $u^{\prime}=\sqrt{2E_{t}/3}$ the root-mean-square velocity, and $E_{t}=\int E(k) \mathrm{d}k$ the total kinetic energy.
The normalized mean enstrophy $\langle\Omega\rangle/(u^{\prime})^2 = \int k^2E(k)\mathrm{d}k/(u^{\prime})^2$ 
is determined by the scale range from $\sigma_{1}$ to $\sigma_{\mathcal{N}}$, since the scale range fully specifies the normalized energy spectrum $E(k)/(u^{\prime})^2$. 
Thus, woven turbulence can be constructed for a specified $Re_{\lambda}$ through determining the scale range from $\sigma_{1}$ to $\sigma_{\mathcal{N}}$.

The model spectrum of turbulence \citep{Pope2000Turbulent}, as an excellent fit to various turbulence data, is specified by
\begin{equation}
\label{eq:model_spectrum}
    E(k)=C_{E}\left(\frac{k \mathcal{L}}{\left[(k \mathcal{L})^2+ 75\right]^{1 / 2}}\right)^{r_{E}-r_{I}}k^{r_{I}}\exp (-4.7 k \eta),
\end{equation}
with $r_{E} = -4/5$, $r_{I} = -5/3$, and the constant $C_{E}$ determined by the total kinetic energy $E_{t}$. 
It is used to assess the spectrum of woven turbulence. 
For example, the spectrum of woven turbulence with $Re_{\lambda}=1237$ agrees with the model spectrum in figure~\ref{fig:Energy spectra}, with smooth transitions between the three wavenumber ranges. 
This agreement confirms that the three-regime energy spectrum is faithfully reproduced in woven turbulence.

\subsection{Intermittency and vortex density}
\label{sec:Intermittency}

Coherent structures in turbulence exhibit intermittency \citep{Buaria2022Vorticity,Buaria2019Extreme,She1990Intermittent}, characterized by sporadic and localized bursts of strong fluctuations of vorticity and dissipation.  
This phenomenon poses a major challenge for turbulence modelling and synthesis. 
In woven turbulence, vorticity is concentrated in vortex tubes that are stochastically distributed in space, leading to local clustering of strong vorticity.
Furthermore, the ratio between regions of intense vorticity and those with zero vorticity can be tuned by adjusting the vortex density $\varrho$ defined in \eqref{eq:vortex density}, thereby influencing the degree of intermittency in woven turbulence.

\begin{figure}
    \centering
   \includegraphics[width=1.0\textwidth]{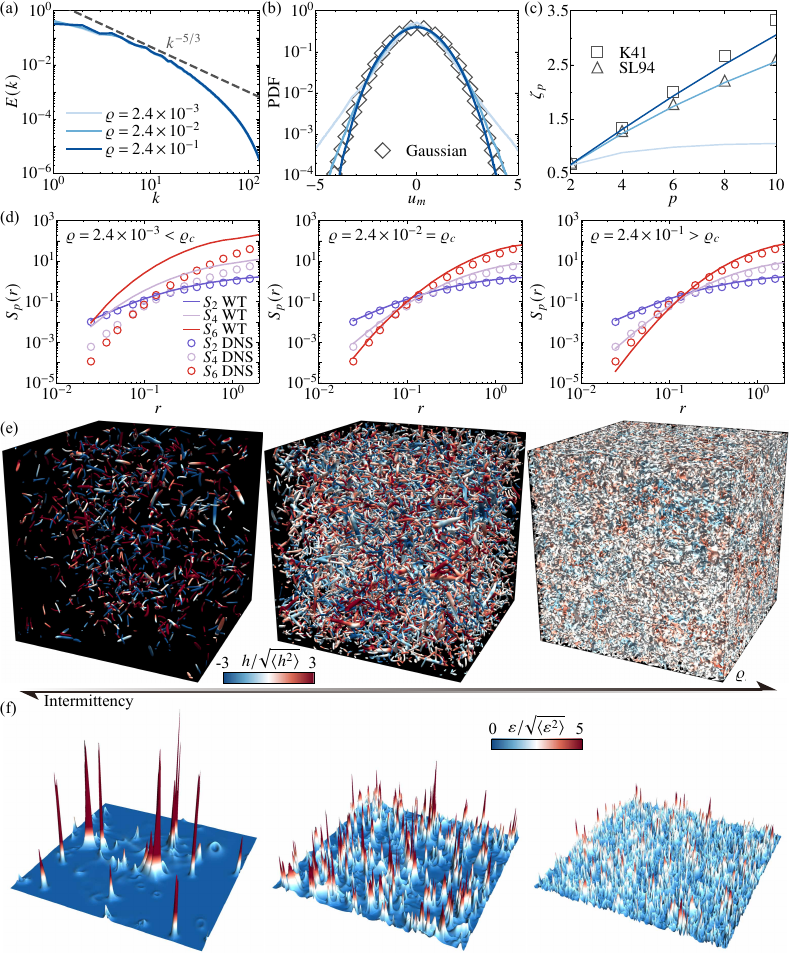}
    \caption{Effect of vortex density $\varrho$ on the intermittency of woven turbulence in cases WT2-I, WT2-C, and WT2-E with varying vortex density $\varrho$ and unity root-mean-square velocity $u^{\prime}$ (see table~\ref{tab:set-up of cases}). 
    (a) Energy spectra with varying $\varrho$. (b) Probability density function (PDF) of velocity components $u_{m}=\boldsymbol{u}\cdot\boldsymbol{e}_{m}$ with varying $\varrho$, where $\boldsymbol{e}_{m}$ is an arbitrary unit vector. (c) Scaling exponents of even-order structure functions with varying $\varrho$, compared with K41 and SL94 model predictions. 
    (d) Second, fourth, and sixth-order structure functions with $\varrho<\varrho_c$, $\varrho=\varrho_c$, and $\varrho>\varrho_c$, along with the corresponding DNS case at $Re_{\lambda}=157$. 
    (e) Isosurfaces of vorticity magnitude $|\boldsymbol{\omega}| = \Gamma_{1}/(4\pi\sigma_{1}^2)$ with $\varrho<\varrho_c$, $\varrho=\varrho_c$, and $\varrho>\varrho_c$ (from left to right), color-coded by the normalized helicity density $h/\sqrt{\langle h^2\rangle}$.
    (f) Distributions of the normalized local energy dissipation $\varepsilon/\sqrt{\langle\varepsilon^2\rangle}$ in the $x$-$y$ slice at $z=0$ with $\varrho<\varrho_c$, $\varrho=\varrho_c$, and $\varrho>\varrho_c$ (from left to right).
    }
    \label{fig:varrho_group}
\end{figure}

To quantify the influence of $\varrho$ on the intermittency in woven turbulence, we construct three test cases: WT2-I with $\varrho = 2.4 \times 10^{-3}$, WT2-C with $\varrho = 2.4 \times 10^{-2}$, and WT2-E with $\varrho = 2.4 \times 10^{-1}$ (see table~\ref{tab:set-up of cases}). 
The value of $\varrho$ is adjusted by varying the centerline length $L_{i}$ of vortex tubes at each scale, while keeping $r_{L}$, $n_{i}$, and $\sigma_{i}$ constant.
Figure~\ref{fig:varrho_group} illustrates vortex structures and representative statistics for these cases.
As $\varrho$ increases, both the vorticity and dissipation fields transition from locally concentrated, highly intermittent structures to nearly uniform, non-intermittent distributions, as illustrated in figures~\ref{fig:varrho_group}(e) and (f), respectively.
At the same time, the energy spectrum remains unchanged in figure~\ref{fig:varrho_group}(a). 
This suggests that the intermittency can be modulated independently of the energy spectrum in woven turbulence.
Such decoupling is consistent with the view that intermittency and spectral scaling represent distinct aspects of turbulence dynamics \citep{Frisch1995Turbulence, Ishihara2009study}.
Furthermore, as $\varrho$ increases, the velocity PDF in figure~\ref{fig:varrho_group}(b) progressively approaches the Gaussian distribution, reflecting the emergence of statistical randomness. 
The presence of statistical randomness, homogeneity, and isotropy in woven turbulence is further analyzed in Appendix~\ref{app:Rand_HIT}. 

\subsubsection{Longitudinal structure functions}
\label{sec:Sp}
Intermittency in turbulence is usually quantified using the scaling exponents of the longitudinal velocity structure functions.
The $p$-th order longitudinal structure function
$S_{p}(r) = \langle[(\boldsymbol{u}(\boldsymbol{x}+r\boldsymbol{e}_{m})-\boldsymbol{u}(\boldsymbol{x}))\cdot\boldsymbol{e}_{m}]^{p}\rangle$ for an arbitrary unit vector $\boldsymbol{e}_{m}$, follows the scaling law
\begin{equation}
    \label{eq:scaling_Sp}
    S_{p}(r) \propto r^{\zeta_{p}}
\end{equation}
with the scaling exponent $\zeta_{p}$ in the inertial range \citep{She1994Universal}.
Unless otherwise specified, structure functions refer to the longitudinal structure functions below.
Intermittency in turbulence can be measured by the deviation of $\zeta_{p}$ from the K41 linear model $\zeta_{p} = p/3$ \citep{Kolmogorov1941The}. 
This deviation is well captured by the SL94 model $\zeta_p = p/9 + 2[1 - (2/3)^{p/3}]$ in \cite{She1994Universal}, reflecting the multifractal characteristics of turbulence. 

\begin{figure}
  \centering
  \includegraphics[width=0.8\textwidth]{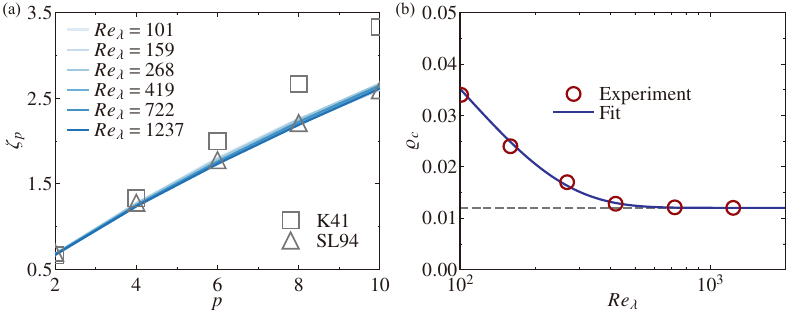}
  \caption{
  Effect of the Taylor-Reynolds number $Re_{\lambda}$ on intermittency and vortex density of woven turbulence.
  (a) Scaling exponents $\zeta_{p}$ of even-order structure functions with varying $Re_{\lambda}$, compared with K41 and SL94 model predictions. (b) Critical vortex density $\varrho_c$ with varying $Re_{\lambda}$.
  }
  \label{fig:varrho_Re}
\end{figure}

In figure~\ref{fig:varrho_group}(c), $\zeta_{p}$ of even-order structure functions deviate significantly from the K41 prediction $\zeta_p = p/3$ at small $\varrho$, reflecting strong intermittency due to sparse and localized vortices in woven turbulence. 
As $\varrho$ increases, $\zeta_{p}$ approaches $p/3$, indicating a diminishing role of intermittency in the limit of large $\varrho$. 
We find that for each $Re_{\lambda}$, there exists a critical vortex density $\varrho = \varrho_c$ such that $\zeta_{p}$ conforms to the SL94 model in figure~\ref{fig:varrho_Re}(a).
In figure~\ref{fig:varrho_Re}(b), the critical vortex density, fitted by
\begin{equation}
\label{eq:critical vortex density}
\varrho_c = 0.07\exp(-Re_{\lambda}/100)+0.012,
\end{equation}
decreases with $Re_{\lambda}$, and converges to a finite value for large $Re_{\lambda}$. 
Meanwhile, the magnitude of the structure functions is also correlated with $\varrho$. 
In figure~\ref{fig:varrho_group}(d), the high-order structure functions are overestimated for $\varrho<\varrho_c$, and they are slightly underestimated at small scales for $\varrho>\varrho_c$.
Note that the odd-order structure functions do not converge (not shown), as the current woven turbulence lacks cross-scale vortex interactions and energy transfer. 
This issue will be addressed in the future work.

The scaling laws of structure functions exhibiting intermittency can be reproduced by most existing fractal models, either monofractal models with decreasing hierarchical vortex density across scales \citep{Frisch1978A} or multifractal models \citep{Meneveau1987Simple}. 
In contrast, our method with a monofractal framework (with constant $r_n = 8$) and scale-invariant vortex density also captures intermittent scaling, where the vortex tubes, as building blocks of woven turbulence, enhances the monofractal framework. 
This emergence of multifractal velocity increments from monofractal vortex tubes highlights the advantage of combining structural and statistical modeling, and suggests that woven turbulence can be further improved using a multifractal framework.

\subsubsection{Vorticity-strain correlation}

The interaction between strain and vorticity is known to drive the formation of extreme events in turbulence \citep{Musci2025Experimental,Zhao2025Evolution}.
This interaction exhibits an asymmetric relationship, which arises from vortex tube dynamics and remains difficult to capture in current statistical models \citep{Buaria2022Vorticity}.
The measurement of this asymmetry is typically based on the local enstrophy $ \Omega  = |\boldsymbol{\omega}|^2/2$ and the squared norm $\Sigma = S_{ij}S_{ij}=\varepsilon/(2\nu)$ of the strain-rate tensor $S_{ij} = \frac{1}{2} \left( {\partial u_i}/{\partial x_j} + {\partial u_j}/{\partial x_i} \right)$. 
In turbulence, local high strain-rate regions consistently exhibit strong vorticity, while intense vorticity leads to only a sublinear increase in the conditional strain. 
These empirical relationships are expressed as \citep{Buaria2022Vorticity}
\begin{equation}
\langle\Omega \mid \Sigma\rangle \sim \Sigma \quad \text{and} \quad \langle\Sigma \mid \Omega\rangle \sim \Omega^\gamma \ (0 < \gamma < 1).
\label{eq:vorticity_strain_relations}
\end{equation}

\begin{figure}
  \centering
  \includegraphics[width=1.0\textwidth]{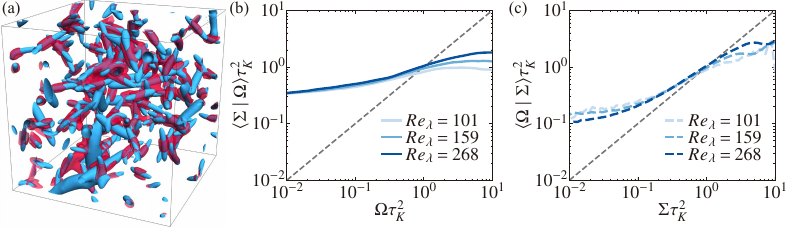}
  \caption{
  Vorticity-strain correlation in woven turbulence.
  (a) Isosurfaces of enstrophy $\Omega=5\langle\Omega\rangle$ (cyan) and squared strain-rate norm $\Sigma=5\langle\Omega\rangle$ (translucent red) in $1/2^3$ spatial portion of case WT2-C. 
  Conditional expectations (b) $\langle\Sigma \mid \Omega\rangle$ and (c) $\langle\Omega \mid \Sigma\rangle$ normalized by Kolmogorov time scale $\tau_{K} = ({\nu}/{\langle \varepsilon \rangle})^{1/2}$ with various $Re_{\lambda}$. 
  Black dashed lines of unit slope are plotted in (b) and (c).
  }
  \label{fig:show_disp_ens}
\end{figure}

Figure~\ref{fig:show_disp_ens}(a) shows the isosurfaces of $\Omega$ and $\Sigma$ in woven turbulence.
The organization of irregular strain structures around strong vortex tubes is similar to that in real turbulence.
This misalignment between $\Omega$ and $\Sigma$ arises from vortex stretching in vortex tubes with varying local core sizes given by \eqref{eq:core size} \citep{Shen2024Designing}. 
Figures~\ref{fig:show_disp_ens}(b) and (c) demonstrate the statistical asymmetry between the enstrophy $\Omega$  and the squared strain-rate norm $\Sigma$ in woven turbulence across different Reynolds numbers, which is qualitatively in agreement with the empirical scaling laws in \eqref{eq:vorticity_strain_relations}. 
Therefore, woven turbulence also captures structural features of extreme events similar to those in real turbulence. 

\section{Fast turbulence synthesis}
\label{sec:Synthetic turbulence}
The DNS of high-Reynolds-number turbulence is extremely costly. 
Instead of solving the NS equations, synthetic turbulence methods generate an instantaneous flow field that mimics turbulence with a low cost. 
The woven turbulence is not only useful for understanding turbulence, but also serves as a fast turbulence synthesis method integrating the coherent vortices and computational efficiency.  
This capability is assessed across a range of $Re_{\lambda}$ from $O(10^2)$ to $O(10^3)$ (see Table~\ref{tab:set-up of cases}) with DNS data (see Table~\ref{tab:DNS cases}), where the methods for generating DNS data are detailed in Appendix~\ref{app:DNS}.

All parameters of woven turbulence cases are uniquely determined by the target DNS cases.
According to the analysis in Sections~\ref{sec:Fractal framework} and \ref{sec:Inertial range}, the self-similar relations \eqref{eq:self-similar} of multi-scale vortex tubes are specified as 
\begin{equation}
\label{eq:self-similar_specified}
    \begin{cases}
    \sigma_{i+1}/\sigma_{i} = 1/2, \\
    L_{i+1}/L_{i} = 1/2, \\
    n_{i+1}/n_{i} = 8, \\
    \Gamma_{i+1}/\Gamma_{i} = (1/2)^{4/3}.
    \end{cases}
\end{equation}
The smallest core size $\sigma_{\mathcal{N}}$ is set according to the matched Kolmogorov length scale $\eta$ by \eqref{eq:scales}. 
Under the self-similar relations \eqref{eq:self-similar_specified} and the scale relations \eqref{eq:scales}, the number of scale levels
\begin{equation}
    \mathcal{N} = \left\lfloor\frac{\mathrm{ln}(\sigma_{\mathcal{N}})-\mathrm{ln}(\mathcal{L}/c_{\mathcal{L}})}{\mathrm{ln}(1/2)}\right\rceil
\end{equation}
is set according to the matched integral scale $\mathcal{L}$.
With the scale range matched, $Re_{\lambda}$ of woven turbulence is thus expected to match that of DNS by \eqref{eq:Re_lambda}. 
The circulation $\Gamma_{i} \propto u^{\prime}$ at each scale is determined by \eqref{eq:self-similar_specified} and the constraint root-mean-square velocity $u^{\prime}=\sqrt{2E_{t}/3}=1$.
The critical vortex density $\varrho_c$ at specified $Re_{\lambda}$ is given by \eqref{eq:critical vortex density}.
Using this value of $\varrho_c$ and the tube population $n_i = 8^{i-1}$, the centerline length $L_i$ at each scale is then determined according to \eqref{eq:self-similar_specified} and \eqref{eq:vortex density}. 
Therefore, the method essentially requires only the Taylor-Reynolds number as input, with no adjustable parameters.

\begin{table}
  \begin{center}
   \setlength{\tabcolsep}{10pt} 
\def~{\hphantom{0}}
  \begin{tabular}{lccc}
       Case & $Re_{\lambda}$ & $\eta$  & $N^3$ \\[3pt]
       DNS1 & 101 & $1.84 \times 10^{-2}$  & $256^3$\\
       DNS2 & 157 & $8.88 \times 10^{-3}$  & $512^3$\\
       DNS3 & 261 & $4.59 \times 10^{-3}$  & $1024^3$\\
       DNS4 & 1257 & $4.98 \times 10^{-4}$  & $8196^3$\\
  \end{tabular}
  \caption{Parameters in DNS cases. 
  }
  \label{tab:DNS cases}
  \end{center}
\end{table}

\subsection{Computational cost}
\label{sec:Numerical performance} 

Since the construction of woven turbulence does not involve temporal evolution, its resolution requirement is weaker than that of DNS.  
We propose the criterion 
\begin{equation}
\label{eq:resolution}
    k_{\rm{max}}\sigma_{\mathcal{N}}>1.5
\end{equation}
for achieving adequate resolution of the smallest scales in woven turbulence with the maximum wavenumber $k_{\rm{max}} \approx N/2$.
This criterion results from resolving more than $95\%$ of the enstrophy of woven turbulence \citep{Pope2000Turbulent}.  
Considering the resolution criterion $k_{\rm{max}}\eta>1.5$ with $k_{\rm{max}} \approx N/3$ for DNS \citep{Pope2000Turbulent} and the relation between $\sigma_{\mathcal{N}}$ and $\eta$ given by \eqref{eq:scales}, the number of grid points required for woven turbulence is $(1/2)^3$ of that needed for DNS with the same Kolmogorov scale.
The resolution of the woven turbulence cases in table~\ref{tab:set-up of cases} satisfies \eqref{eq:resolution} for resolving the smallest scales.
Moreover, since the woven turbulence is constructed directly without temporal evolution, if the resolution criterion \eqref{eq:resolution} is not satisfied, the flow field within the resolved wavenumber range will not be affected.

We have systematically redesigned the numerical algorithm in \citet{Shen2024Designing} to significantly enhance the computational efficiency for generating woven turbulence.
As discussed in Appendix~\ref{app:Numerical}, the cost for constructing woven turbulence scales as $O(\varrho_c N^3)$. 
Since $\varrho_c$ approaches a constant at high Reynolds numbers in \eqref{eq:critical vortex density}, the effective scaling is $O(N^3)$, which is the optimal scaling for 3D turbulence synthesis.
In contrast, the computational cost of the random Fourier modes is $O(N^{3}\log N)$ \citep{Fung1992Kinematic}, consistent with that of the 3D fast Fourier transform.
Most existing fractal models exhibit the same asymptotic complexity \citep{Malara2016Fast}, where the logarithmic factor $\log N$ arises from the increasing total vortex density with Reynolds number.
Given $N\propto Re_{\lambda}^{3/2}$, the computational cost for constructing woven turbulence is $O(Re_{\lambda}^{9/2})$. 
This is significantly less than $O(Re_{\lambda}^{6})$ required by DNS \citep{Pope2000Turbulent}, which is obtained after running multiple time steps until reaching a statistically stationary state to obtain an instantaneous turbulent field. 

\begin{figure}
    \centering
    \includegraphics[width=1.0\textwidth]{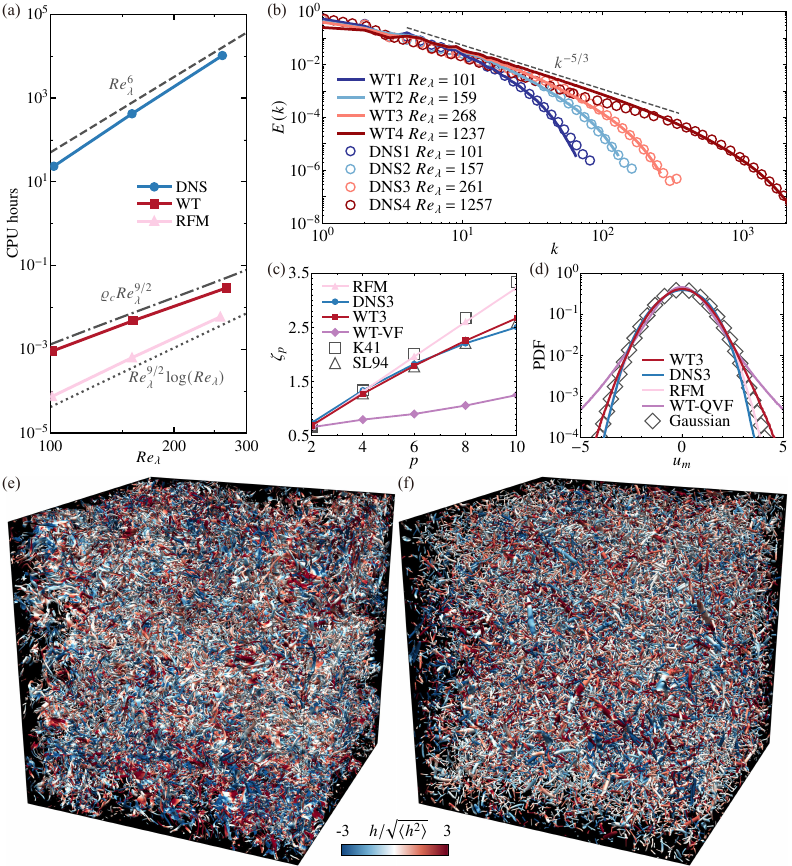}
    \caption{Comparison between woven turbulence cases in table~\ref{tab:set-up of cases} and DNS cases in table~\ref{tab:DNS cases}.
    (a) Computational costs with varying $Re_{\lambda}$. 
    Here, the energy spectra of the random Fourier modes (RFM) are set to be the same as those of woven turbulence at the same $Re_{\lambda}$.
    (b) Energy spectra with various $Re_{\lambda}$. 
    (c) Scaling exponents $\zeta_{p}$ of structure functions with even $p$. 
    The setting of the woven turbulence based on vortex filaments (WT-VF) is the same as the case with $Re_{\lambda}=161.4$ in \citet{Shen2024Designing}.
    (d) PDFs of velocity components.
    Isosurfaces of $|\boldsymbol{\omega}|=3\sqrt{\langle\Omega\rangle}$ for (e) case DNS3 with the computational cost $1.0\times 10^{4}$ CPU hours and (f) case WT3 with the computational cost $2.9\times 10^{-2}$ CPU hours. These isosurfaces are color-coded by $h/\sqrt{\langle h^2\rangle}$.}
    \label{fig:case_group}
\end{figure}

Figure~\ref{fig:case_group}(a) compares the computational costs of generating an instantaneous turbulent flow field using DNS, woven turbulence, and random Fourier modes at various $Re_{\lambda}$. 
These simulations were performed on the TianheXY-C  at the National Supercomputer Center in Guangzhou, China. 
For $Re_{\lambda}>200$, the computational cost of woven turbulence is less than $10^{-5}$ of that of DNS, and this difference becomes more significant as $Re_{\lambda}$ increases. 
Moreover, although the cost of random Fourier modes is lower, the gap with woven turbulence narrows as $Re_\lambda$ increases. 
Therefore, our method exhibits high computational efficiency in generating instantaneous turbulent fields.

\subsection{Assessment of woven turbulence} 
\label{sec:Validation}
The quality of woven turbulence is assessed by comparing its flow statistics and vortex structures with those of DNS at a range of $Re_\lambda$ from $O(10^2)$ to $O(10^3)$. 
Figure~\ref{fig:case_group}(b) shows an excellent agreement of the energy spectra of DNS and woven turbulence. Note that both are normalized by setting
the total energy $E_{t}=3/2$ for $u^{\prime}=1$. 
For the woven turbulence, $E(k)$ in the inertial range follows Kolmogorov's $-5/3$ law, and the inertial range expands with increasing $Re_\lambda$; $E(k)$ in the dissipation range exhibits exponential decay with a smooth transition to that in the inertial range. 

In figure~\ref{fig:case_group}(c), scaling exponents $\zeta_{p}$ of structure functions in woven turbulence also closely resemble those of DNS, and both align well with the SL94 model. 
This indicates that the woven turbulence exhibits intermittency similar to real turbulence, which is one of the key distinctions between turbulent fields and Gaussian random fields.
As a comparison, for the woven turbulence based on vortex filaments (WT-VF) in \cite{Shen2024Designing}, $\zeta_{p}$ grows slowly with $p$, which indicates excessively strong intermittency. 
In contrast, the random Fourier modes cannot reproduce intermittency, and its $\zeta_{p}$ conforms to the K41 linear prediction.
Additionally, in figure~\ref{fig:case_group}(d), the velocity PDF of woven turbulence exhibits the Gaussian distribution, consistent with that of DNS.

Figures~\ref{fig:case_group}(e) and (f) illustrate the vortex structures in DNS and woven turbulence, respectively. 
Both exhibit elongated, intertwined vortices forming complex coherent structures with similar sizes and spatial distributions.
The generation of such coherent vortices similar to real turbulence is unique among turbulence synthesis methods.
On the other hand, the axisymmetric kernel function \eqref{eq:kernel} gives rise to tubular vortices in woven turbulence, whereas those in DNS exhibit both tube- and sheet-like structures. 
The incorporation of sheet-like vortices \citep{Shen2024Designing} into the woven turbulence can be considered in the future work. 

\section{Conclusions}
\label{sec:conclusion}

We model turbulence using multi-scale coherent vortices, explicitly constructed along stochastic vortex centerlines produced by FBBs and organized within a statistical fractal framework. 
The parameters of these vortex tubes are scale-dependent but uniform within each scale, and the scaling exponents are determined by dimensional analysis and geometric similarity.
By combining the strengths of structural and statistical modelling, this approach naturally captures key turbulence features, including the three-regime energy spectrum, intermittency, and coherent structures.

By varying key parameters of vortex tubes, we quantify several relations of statistics and structures in woven turbulence. 
First, the scale-invariance of the hierarchical vortex density $\varrho_{i}$ corresponds to Kolmogorov's $-5/3$ law in the inertial range.
Second, as the total vortex density $\varrho$ increases, the woven turbulence transitions from a state of vortex clustering with excessive intermittency to an approximately uniform Gaussian random field without intermittency.
For each $Re_{\lambda}$, there exists a critical vortex density $\varrho_c$ such that woven turbulence captures intermittent scaling laws of structure functions and the asymmetric vorticity-strain correlation comparable to real turbulence.
This critical vortex density decreases with $Re_{\lambda}$ and converges to a finite value in the inviscid limit.
Furthermore, the spatial arrangement of vortex tubes influences the energy-containing range, and their internal structure affects the dissipation range, where the energy spectrum exhibits exponential decay similar to real turbulence.

Our method also enables fast synthesis of turbulence at specified Reynolds numbers without adjustable parameters.
The statistical features and coherent vortices of the woven turbulence at various $Re_{\lambda}$ are in good agreement with the corresponding DNS results. 
In particular, the computational cost of woven turbulence achieves the optimal scaling $O(N^3)$ for 3D turbulence synthesis.
For $Re_{\lambda}>200$, the computational cost of generating an instantaneous turbulent field using woven turbulence is less than $10^{-5}$ that of DNS, and this difference becomes more significant as $Re_{\lambda}$ increases.
Thus our method is suitable for rapidly generating massive turbulence data at high Reynolds numbers with controllable statistical and structural features, for applications such as turbulence simulations, machine learning training, and turbulence model assessment.

Overall, this work proposes a computationally efficient turbulence modeling approach that integrates the explicit construction of coherent vortices with a multi-scale statistical framework. 
However, the internal structure of vortices and their spatial arrangements is still inadequate, limiting the model's ability to reproduce key asymmetric statistics of real turbulence.
For instance, both the PDF of longitudinal velocity increments and the joint PDF of the second ($Q$) and third ($R$) invariants of the velocity-gradient tensor exhibit pronounced asymmetry in real turbulence, whereas the present model yields symmetric distributions.

In the future work, the present approach can be improved in several aspects. 
First, incorporating multi-scale interaction into centerline generation is expected to reproduce energy transfer across scales \citep{Shen2024Designing}.
Notably, energy transfer across scales corresponds to non-trivial third-order structure functions, thereby linking to the asymmetric PDF of longitudinal velocity increments \citep{Frisch1995Turbulence}.
Second, refining the internal structure of vortices can better capture fine-scale nonlinear dynamics. 
For example, employing Lundgren vortices \citep{Lundgren1982Strained} that exhibit strong vortex stretching and spiraling may enable reproduction of the asymmetric local geometry represented by the $Q$-$R$ joint PDF. 
Other structures, such as vortex sheets, could also be included to increase structural diversity. 
Third, multifractal modelling \citep{Meneveau1987Simple,Malara2016Fast,Zhou2015Multifractal,L2023Stochastic} will be explored to capture finer details of intermittency beyond what is achieved with the current method.

\backsection[Acknowledgments]{Numerical simulations were carried out on the TianheXY-C supercomputer in Guangzhou, China.}
\backsection[Funding]{This work has been supported by the National Natural Science Foundation of China (Grant Nos.~12432010, 12525201 and 12588201), and the Xplore Prize.}
\backsection[Declaration of interests]{The authors report no conflict of interest.}
\backsection[Author contributions]{Y.Y. and Z.H. designed research. Z.H. performed research. All the authors discussed the results and wrote the manuscript. All the authors have given approval for the manuscript.}
\backsection[Data availability statement]{
The code that support the findings of this study is openly available at https://github.com/YYgroup/FastWeavTurb.
}

\appendix

\section{Numerical construction of woven turbulence}
\label{app:Numerical}

\subsection{Vortex centerline construction}
\label{app:numerical_centerline}

\begin{figure}
  \centering
  \includegraphics[width=1.0\textwidth]{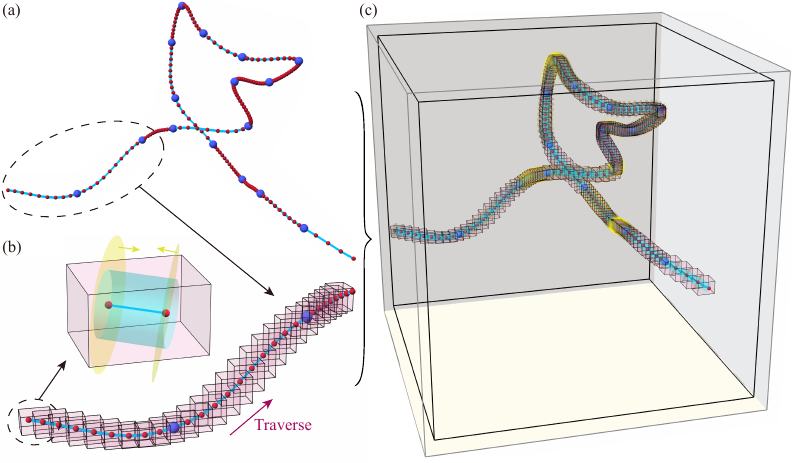}
  \caption{
  Schematic of the vortex tube construction algorithm in woven turbulence.
  This illustration is based on a segment of a vortex tube from the woven turbulence case WT2 (see table~\ref{tab:set-up of cases}).
  (a) Vortex centerline construction and division. The twice continuously differentiable vortex centerline (cyan curve) is interpolated from the FBB points (blue spheres) as in \eqref{eq:spline}. For further local vorticity calculation, the centerline is dynamically divided (red spheres) into sub-segments based on its local bending, as specified in \eqref{eq:dividing points} and \eqref{eq:NC}. 
  (b) Region for calculating local vorticity in the vortex tube. 
  First, search regions (transparent pink boxes) are sequentially established along the centerline based on the sub-segments, as given in \eqref{eq:box}. Then, a cylindrical-like region is extracted within each search region (transparent cyan cylinders bounded by two directional yellow disks), as described by \eqref{eq:subdomain}. 
  Finally, the local vorticity field is computed within each of these cylindrical-like regions.
  (c) Integrated visualization of the vortex tube construction in woven turbulence.
  The black-bordered box represents the flow field domain $V$, while the surrounding gray-bordered box indicates the ghost-point extension of $V$ for implementing periodic boundary conditions. 
  }
  \label{fig:algorithm}
\end{figure}

The woven turbulence construction algorithm is illustrated in figure~\ref{fig:algorithm}.
This algorithm is a systematic reformulation of that in \citet{Shen2024Designing}, achieving significantly higher computational efficiency and modestly improved accuracy.

For constructing vortex centerlines, the FBB is constructed by wavelet-based synthesis \citep{Abry1996the,Bardet2003theory}. 
Compared to the construction of vortex tubes based on given centerlines, the cost of generating the FBB is negligible. 
We apply the fifth-order spline interpolation on the FBB points to obtain a twice continuously differentiable vortex centerline.
The blue spheres and the cyan curve in figure~\ref{fig:algorithm}(a) illustrate the FBB points and the interpolated centerline, respectively.
First, we treat the FBB points as a sequence of discrete control points ${\boldsymbol{c}}_{J}=\left(x_{J}, y_{J}, z_{J}\right)=\boldsymbol{B}(J)$ on the centerline with $J=1,2, \ldots, \mathscr{N}$, and introduce the cumulative chord length
\begin{equation}
\label{eq:disc_para}
\mathscr{S}_{J}=
\begin{cases}
0, & {J}=1, \\
\sum_{{J}^{\prime}=2}^{J}\left|{\boldsymbol{c}}_{{J}^{\prime}}-{\boldsymbol{c}}_{{J}^{\prime}-1}\right|, & {J}=2,3, \ldots, \mathscr{N}, \\
\left|{\boldsymbol{c}}_{1}-{\boldsymbol{c}}_{\mathscr{N}}\right|+\sum_{{J}=2}^{\mathscr{N}}\left|{\boldsymbol{c}}_{J}-{\boldsymbol{c}}_{{J}-1}\right|, & {J}=\mathscr{N}+1
\end{cases}
\end{equation}
with the total cumulative chord length $\widetilde{L}=\mathscr{S}_{\mathscr{N}+1}$.
Then we obtain the normalized cumulative chord length 
\begin{equation}
\label{eq:norm_disc_para}
{\mathcal{S}}_{J}=\frac{2 \pi \mathscr{S}_{J}}{\widetilde{L}} \in[0, 2\pi], \quad {J}=1,2, \ldots, \mathscr{N}+1.
\end{equation}

Next, we employ the fifth-order spline parametric equation
\begin{equation}
\label{eq:spline}
\boldsymbol{c}(\mathcal{S})=\boldsymbol{A}_{J} \mathcal{S}^5+\boldsymbol{B}_{J} \mathcal{S}^4+\boldsymbol{C}_{J} \mathcal{S}^3+\boldsymbol{D}_{J} \mathcal{S}^2+\boldsymbol{E}_{J} \mathcal{S}+\boldsymbol{F}_{J},
\quad \mathcal{S} \in\left[{\mathcal{S}}_{J}, {\mathcal{S}}_{{J}+1}\right)
\end{equation}
to smoothly connect the control points in a piecewise manner, where $\mathcal{S}$ is the spline parameter based on the normalized cumulative chord length $\mathcal{S}_{J}$ in \eqref{eq:norm_disc_para}.
Specifically, in each segment, we determine six vector coefficients
$[\boldsymbol{A}_{J},\boldsymbol{B}_{J},\boldsymbol{C}_{J},\boldsymbol{D}_{J},\boldsymbol{E}_{J},\boldsymbol{F}_{J}]$
by solving the linear system
\begin{equation}
\label{eq:spline parameters}
    \begin{bmatrix}
    \boldsymbol{c}({\mathcal{S}}_{J})\\\boldsymbol{c}({\mathcal{S}}_{{J}+1})\\\frac{\mathrm{d}\boldsymbol{c}}{\mathrm{d}\mathcal{S}}({\mathcal{S}}_{J})\\\frac{\mathrm{d}\boldsymbol{c}}{\mathrm{d}\mathcal{S}}({\mathcal{S}}_{{J}+1})\\\frac{\mathrm{d^2}\boldsymbol{c}}{\mathrm{d}\mathcal{S}^2}({\mathcal{S}}_{J})\\\frac{\mathrm{d^2}\boldsymbol{c}}{\mathrm{d}\mathcal{S}^2}({\mathcal{S}}_{{J}+1})\end{bmatrix}=\begin{bmatrix}{\mathcal{S}}_{J}^5&{\mathcal{S}}_{J}^4&{\mathcal{S}}_{J}^3&{\mathcal{S}}_{J}^2&{\mathcal{S}}_{J}&1\\{\mathcal{S}}_{{J}+1}^5&{\mathcal{S}}_{{J}+1}^4&{\mathcal{S}}_{{J}+1}^3&{\mathcal{S}}_{{J}+1}^2&{\mathcal{S}}_{{J}+1}&1\\5{\mathcal{S}}_{J}^4&4{\mathcal{S}}_{J}^3&3{\mathcal{S}}_{J}^2&2{\mathcal{S}}_{J}&1&0\\5{\mathcal{S}}_{{J}+1}^4&4{\mathcal{S}}_{{J}+1}^3&3{\mathcal{S}}_{{J}+1}^2&2{\mathcal{S}}_{{J}+1}&1&0\\20{\mathcal{S}}_{J}^3&12{\mathcal{S}}_{J}^2&6{\mathcal{S}}_{J}&2&0&0\\20{\mathcal{S}}_{{J}+1}^3&12{\mathcal{S}}_{{J}+1}^2&6{\mathcal{S}}_{{J}+1}&2&0&0\end{bmatrix}\begin{bmatrix}\boldsymbol{A}_{J}\\\boldsymbol{B}_{J}\\\boldsymbol{C}_{J}\\\boldsymbol{D}_{J}\\\boldsymbol{E}_{J}\\\boldsymbol{F}_{J}\end{bmatrix}
\end{equation}
with
\begin{equation}
    \label{eq:dspline}
    \frac{\mathrm{d}\boldsymbol{c}}{\mathrm{d}\mathcal{S}}({\mathcal{S}})=5\boldsymbol{A}_{J} \mathcal{S}^4+4\boldsymbol{B}_{J} \mathcal{S}^3+3\boldsymbol{C}_{J} \mathcal{S}^2+2\boldsymbol{D}_{J} \mathcal{S}+\boldsymbol{E}_{J},
    \quad \mathcal{S} \in\left[{\mathcal{S}}_{J}, {\mathcal{S}}_{{J}+1}\right)
\end{equation}
and
\begin{equation}
    \label{eq:ddspline}
    \frac{\mathrm{d}^2\boldsymbol{c}}{\mathrm{d}\mathcal{S}^2}({\mathcal{S}})=20\boldsymbol{A}_{J} \mathcal{S}^3+12\boldsymbol{B}_{J} \mathcal{S}^2+6\boldsymbol{C}_{J} \mathcal{S}+2\boldsymbol{D}_{J},
    \quad \mathcal{S} \in\left[{\mathcal{S}}_{J}, {\mathcal{S}}_{{J}+1}\right).
\end{equation}
The fifth-order spline interpolation ensures second-order derivative continuity, thereby guaranteeing the continuity of the Frenet frame
\begin{equation}
\label{eq:Frenet}
\begin{cases}
{\boldsymbol{T}}(\mathcal{S}) = \frac{\mathrm{d}\boldsymbol{c}}{\mathrm{d}\mathcal{S}}({\mathcal{S}})/\left|\frac{\mathrm{d}\boldsymbol{c}}{\mathrm{d}\mathcal{S}}({\mathcal{S}})\right|, \\
{\boldsymbol{N}}(\mathcal{S}) = \frac{\mathrm{d}\boldsymbol{T}}{\mathrm{d}\mathcal{S}}({\mathcal{S}})/\left|\frac{\mathrm{d}\boldsymbol{T}}{\mathrm{d}\mathcal{S}}({\mathcal{S}})\right|, \\
{\boldsymbol{B}}(\mathcal{S}) = {\boldsymbol{T}}(\mathcal{S})\times{\boldsymbol{N}}(\mathcal{S})
\end{cases}
\end{equation}
and the Frenet frame satisfies
\begin{equation}
    \begin{cases} \frac{\mathrm{d} \boldsymbol{T}}{\mathrm{d} s} =\kappa \boldsymbol{N}, \\
    \frac{\mathrm{d} \boldsymbol{N}}{\mathrm{d} s} =-\kappa \boldsymbol{T}+\tau \boldsymbol{B}, \\
    \frac{\mathrm{d} \boldsymbol{B}}{\mathrm{d} s} =-\tau \boldsymbol{N}
    \end{cases}
\end{equation}
with $\kappa$ denotes the curvature and $\tau$ the torsion.
The Frenet frame is applied in the subsequent construction of the vorticity field, and its continuity is the basis for the continuity of vorticity.
Finally, the spline given by \eqref{eq:spline} is mapped into the periodic box ${V}$ by taking the modulo of its coordinates with respect to the box side length $\mathscr{L}$.

Furthermore, the arc-length of the centerline is
\begin{equation}
\label{eq:arc-length}
s(\mathcal{S})= \int_{0}^{\mathcal{S}}\left|\frac{\mathrm{d}\boldsymbol{c}(\mathcal{S}^{\prime})}{\mathrm{d}\mathcal{S}^{\prime} }\right|\mathrm{d}\mathcal{S}^{\prime}.
\end{equation}
Since there is a one-to-one mapping between $s$ and $\mathcal{S}$, the points on the centerline can be re-parameterized by $s$, i.e., $\boldsymbol{c}(s)=\boldsymbol{c}(s(\mathcal{S}))$.
In addition, the length 
\begin{equation}
    L = \int_{\mathcal{C}} \mathrm{d}s \approx c_{H} \delta_B \mathscr{N}
\end{equation}
of centerline $\mathcal{C}$ is proportional to $\mathscr{N}$ with $c_{H} = 1.2$ for $H=5/6$.

\subsection{Vorticity field construction}
The vorticity field \eqref{eq:w_sum} is generated on a Cartesian grid with $N^3$ uniform grid points. 
We construct the local vorticity field along each vortex centerline as shown in figure~\ref{fig:algorithm}(b), instead of determining which vortex tube each grid point belongs to as in \citet{Shen2024Designing}.
To this end, a fine discretization of the given segmented parametric centerline $\mathcal{C}:\boldsymbol{c}(\mathcal{S})$ with $\mathcal{S} \in [0,2\pi)$ is required.
Specifically, we divide each segment of $\mathcal{C}$ into $N_{C}$ uniform sub-segments by $N_{C}$ dividing points
\begin{equation}
\label{eq:dividing points}
\boldsymbol{c}_{{J},{K}} = \boldsymbol{c} \left(\mathcal{S}_{{J},{K}}\right), \quad {J}=1,2,\ldots,\mathscr{N},
\quad {K}=1,2,\ldots,N_{C}
\end{equation}
with $\mathcal{S}_{{J},{K}} = {\mathcal{S}}_{J} + ({K}-1)({\mathcal{S}}_{{J}+1}-{\mathcal{S}}_{J})/N_{C}$.
The red spheres in figure~\ref{fig:algorithm}(a) illustrate these dividing points.
To balance computational accuracy and efficiency, the number of subdivisions
\begin{equation}
\label{eq:NC}
    N_{C} = M_{C}  \int_{\mathcal{S}_{J}}^{\mathcal{S}_{{J}+1}}\kappa(\mathcal{S})\left|\frac{\mathrm{d}\boldsymbol{c}}{\mathrm{d}\mathcal{S} }\right|\mathrm{d}\mathcal{S}
\end{equation}
is dynamically determined based on the local bending of the centerline, where $M_{C}=50$ is a resolution parameter calibrated such that a perfect circle is divided into $M_{C}$ uniform sub-segments and the curvature is calculated as
\begin{equation}
    \kappa(\mathcal{S}) = {\left|\frac{\rm{d}\boldsymbol{c}}{\rm{d}\mathcal{S}}\times\frac{\rm{d}^2\boldsymbol{c}}{\rm{d}\mathcal{S}^2}\right|}\bigg/{\left|\frac{\rm{d}\boldsymbol{c}}{\rm{d}\mathcal{S}}\right|^3}.
\end{equation}

Based on these dividing points, the space in the proximity of centerline $\mathcal{C}$ can be divided into several subdomains
\begin{equation}
\label{eq:subdomain}
{V}_{{J},{K}} = \left\{\boldsymbol{x} \middle|
\begin{aligned}
    &
    \left(\boldsymbol{x} - \boldsymbol{c}_{{J},{K}}\right) \cdot \boldsymbol{T}_{{J},{K}} \geqslant 0 , \left(\boldsymbol{x} - \boldsymbol{c}_{{J},{K}+1}\right) \cdot \boldsymbol{T}_{{J},{K}+1} < 0
    \\
    &\text{ and } 
    \left|\left(\boldsymbol{x} - \boldsymbol{c}_{{J},{K}}\right)\times
    \left(\boldsymbol{c}_{{J},{K}+1} - \boldsymbol{c}_{{J},{K}}\right)\right|/
    \left|\boldsymbol{c}_{{J},{K}+1} - \boldsymbol{c}_{{J},{K}}\right|<\mathcal{R}
\end{aligned}
\right\},
\end{equation}
with 
$
\boldsymbol{T}_{{J},{K}} = \boldsymbol{T} \left(\mathcal{S}_{{J},{K}}\right),
$
where the subscript ``${{J},N_{C}+1}$'' is equivalent to ``${{J}+1,1}$''. 
The local vorticity field is calculated in these subdomains.
In figure~\ref{fig:algorithm}(b), the transparent cyan tube represents the condition 
\begin{equation}
\label{eq:subdomain_c1}
    \left|\left(\boldsymbol{x} - \boldsymbol{c}_{{J},{K}}\right)\times
\left(\boldsymbol{c}_{{J},{K}+1} - \boldsymbol{c}_{{J},{K}}\right)\right|/
\left|\boldsymbol{c}_{{J},{K}+1} - \boldsymbol{c}_{{J},{K}}\right|<\mathcal{R}
\end{equation}
in \eqref{eq:subdomain}, which is used to determine whether a spatial point lies within the vortex tube.
The two transparent yellow oriented disks in figure~\ref{fig:algorithm}(b) represent the conditions
\begin{equation}
\label{eq:subdomain_c2}
    \left(\boldsymbol{x} - \boldsymbol{c}_{{J},{K}}\right) \cdot \boldsymbol{T}_{{J},{K}} \geqslant 0\quad \textrm{and} \quad \left(\boldsymbol{x} - \boldsymbol{c}_{{J},{K}+1}\right) \cdot \boldsymbol{T}_{{J},{K}+1} < 0
\end{equation}
in \eqref{eq:subdomain}, respectively, which together serve to distinguish different subdomains along the vortex tube. 
The intersection of the regions defined by \eqref{eq:subdomain_c1} and \eqref{eq:subdomain_c2} thus defines the subdomain ${V}_{{J},{K}}$ in \eqref{eq:subdomain}. 
The transparent cyan tubular region in figure~\ref{fig:algorithm}(c) illustrates the curved tubular structure formed by multiple connected ${V}_{{J},{K}}$ subdomains.
We set $\mathcal{R}=3\sigma(s)$ so that the vortex tube contains over $99.7\%$ of the vorticity in \eqref{eq:w_single}.

To optimize computational efficiency, the search for subdomains ${V}_{{J},{K}}$ and subsequent vorticity calculations are performed within local bounding boxes
\begin{equation}
\label{eq:box}
\begin{cases}
i_{x,\mathrm{min}} = \left\lfloor \dfrac{ \min( \boldsymbol{c}_{J,K} \cdot \boldsymbol{e}_x,\, \boldsymbol{c}_{J,K+1} \cdot \boldsymbol{e}_x ) - \mathcal{R} }{ \mathscr{L} } N \right\rfloor, 
\\[1em]
i_{x,\mathrm{max}} = \left\lceil \dfrac{ \max( \boldsymbol{c}_{J,K} \cdot \boldsymbol{e}_x,\, \boldsymbol{c}_{J,K+1} \cdot \boldsymbol{e}_x ) + \mathcal{R} }{ \mathscr{L} } N \right\rceil, 
\\[1em]
i_{y,\mathrm{min}} = \left\lfloor \dfrac{ \min( \boldsymbol{c}_{J,K} \cdot \boldsymbol{e}_y,\, \boldsymbol{c}_{J,K+1} \cdot \boldsymbol{e}_y ) - \mathcal{R} }{ \mathscr{L} } N \right\rfloor, 
\\[1em]
i_{y,\mathrm{max}} = \left\lceil \dfrac{ \max( \boldsymbol{c}_{J,K} \cdot \boldsymbol{e}_y,\, \boldsymbol{c}_{J,K+1} \cdot \boldsymbol{e}_y ) + \mathcal{R} }{ \mathscr{L} } N \right\rceil, 
\\[1em]
i_{z,\mathrm{min}} = \left\lfloor \dfrac{ \min( \boldsymbol{c}_{J,K} \cdot \boldsymbol{e}_z,\, \boldsymbol{c}_{J,K+1} \cdot \boldsymbol{e}_z ) - \mathcal{R} }{ \mathscr{L} } N \right\rfloor, 
\\[1em]
i_{z,\mathrm{max}} = \left\lceil \dfrac{ \max( \boldsymbol{c}_{J,K} \cdot \boldsymbol{e}_z,\, \boldsymbol{c}_{J,K+1} \cdot \boldsymbol{e}_z ) + \mathcal{R} }{ \mathscr{L} } N \right\rceil, 
\end{cases}
\end{equation}
shown as transparent pink boxes in figure~\ref{fig:algorithm}(b).
These local bounding boxes are swept along the centerline and may extend beyond the flow field domain $V$ into the ghost-point extension (shown as gray boxes in figure~\ref{fig:algorithm}(c)) designed to facilitate periodic boundary condition implementation.
Since the total volume of these boxes is statistically proportional to the number of grid points within vortex tubes, and this number is proportional to the vortex density $\varrho$ defined in \eqref{eq:vortex density}, the computational cost consequently scales as $O(\varrho N^3)$.

Within each identified subdomain ${V}_{{J},{K}}$, the distance from a point $\boldsymbol{x}$ to the centerline $\mathcal{C}$ is given by
\begin{equation}
\label{eq:rho_x}
\widetilde{\rho}_{\boldsymbol{x}} = \left|\boldsymbol{x} - \boldsymbol{c} \left(\widetilde{\mathcal{S}}_{\boldsymbol{x}}\right)\right|,
\end{equation}
where
\begin{equation}
\label{eq:zeta_x}
\widetilde{\mathcal{S}}_{\boldsymbol{x}} = \frac{\mathcal{S}_{{J},{K}+1} \left(\boldsymbol{x} - \boldsymbol{c}_{{J},{K}}\right) \cdot \boldsymbol{T}_{{J},{K}} + \mathcal{S}_{{J},{K}} \left(\boldsymbol{c}_{{J},{K}+1} - \boldsymbol{x}\right) \cdot \boldsymbol{T}_{{J},{K}}}{\left|\boldsymbol{c}_{{J},{K}+1} - \boldsymbol{c}_{{J},{K}}\right|}
\end{equation}
is the spline parameter corresponding to the nearest point on $\mathcal{C}$ to $\boldsymbol{x}$. 
The azimuth-related functions and arc-length corresponding to $\boldsymbol{x}$ are calculated by
\begin{equation}
\begin{cases}
\cos\widetilde{\theta}_{\boldsymbol{x}} = \frac{\left(\boldsymbol{x} - \boldsymbol{c} \left(\widetilde{\mathcal{S}}_{\boldsymbol{x}}\right)\right) \cdot \boldsymbol{N}\left(\widetilde{\mathcal{S}}_{\boldsymbol{x}}\right)}{\widetilde{\rho}_{\boldsymbol{x}}}, \\[0.5ex]
\sin\widetilde{\theta}_{\boldsymbol{x}} = \frac{\left(\boldsymbol{x} - \boldsymbol{c} \left(\widetilde{\mathcal{S}}_{\boldsymbol{x}}\right)\right) \cdot {\boldsymbol{B}}\left(\widetilde{\mathcal{S}}_{\boldsymbol{x}}\right)}{\widetilde{\rho}_{\boldsymbol{x}}}
\end{cases}
\end{equation}
and
$\widetilde{s}_{\boldsymbol{x}} = s(\widetilde{\mathcal{S}}_{\boldsymbol{x}})$, respectively.
Thus the local cylindrical frame becomes
\begin{equation}
\label{eq:LCF_x}
\begin{cases}
\widetilde{\boldsymbol{e}}_{s,\boldsymbol{x}} = {\boldsymbol{T}}\left(\widetilde{\mathcal{S}}_{\boldsymbol{x}}\right), \\
\widetilde{\boldsymbol{e}}_{\rho,\boldsymbol{x}} = \cos\widetilde{\theta}_{\boldsymbol{x}} {\boldsymbol{N}}\left(\widetilde{\mathcal{S}}_{\boldsymbol{x}}\right) + \sin\widetilde{\theta}_{\boldsymbol{x}} {\boldsymbol{B}}\left(\widetilde{\mathcal{S}}_{\boldsymbol{x}}\right), \\
\widetilde{\boldsymbol{e}}_{\theta,\boldsymbol{x}} = -\sin\widetilde{\theta}_{\boldsymbol{x}} {\boldsymbol{N}}\left(\widetilde{\mathcal{S}}_{\boldsymbol{x}}\right) + \cos\widetilde{\theta}_{\boldsymbol{x}} {\boldsymbol{B}}\left(\widetilde{\mathcal{S}}_{\boldsymbol{x}}\right).
\end{cases}
\end{equation}
Then we approximate \eqref{eq:w_single} as
\begin{equation}
\label{eq:w_x}
\widetilde{\boldsymbol{\omega}}_{ij}(\boldsymbol{x})= 
\begin{cases}
\Gamma_{i} G_{i} \left(\widetilde{s}_{\boldsymbol{x}}, \widetilde{\rho}_{\boldsymbol{x}}\right) \left[\widetilde{\boldsymbol{e}}_{s,\boldsymbol{x}} + \frac{\widetilde{\rho}_{\boldsymbol{x}}}{R_{i} \left(\widetilde{s}_{\boldsymbol{x}}\right) \left(1 - \widetilde{\kappa}_{\boldsymbol{x}} \widetilde{\rho}_{\boldsymbol{x}} \cos\widetilde{\theta}_{\boldsymbol{x}}\right)} \frac{{\rm d}R_{i}}{{\rm d}s} \left(\widetilde{s}_{\boldsymbol{x}}\right)
\widetilde{\boldsymbol{e}}_{\rho,\boldsymbol{x}}  \right], & 1 > \kappa \left(\widetilde{\mathcal{S}}_{\boldsymbol{x}}\right) \widetilde{\rho}_{\boldsymbol{x}} \cos\widetilde{\theta}_{\boldsymbol{x}} \\
\boldsymbol{0}, & 1 \leqslant \kappa \left(\widetilde{\mathcal{S}}_{\boldsymbol{x}}\right)\widetilde{\rho}_{\boldsymbol{x}} \cos\widetilde{\theta}_{\boldsymbol{x}}
\end{cases}
\end{equation}
with computed and given variables. 

The above describes the construction of a single vortex tube as shown in figure~\ref{fig:algorithm}(c).
For multi-scale vortex tubes, the total vorticity field is obtained by summing the individual vorticity fields of $\sum_{i=1}^{\mathcal{N}} n_i$ vortex tubes, each computed via \eqref{eq:w_x}, according to the superposition rule given in \eqref{eq:w_sum}.
The procedure for computing $\boldsymbol{\omega}(\boldsymbol{x})$ in woven turbulence is outlined in Algorithm~\ref{algorithm}.
The corresponding code is available at https://github.com/YYgroup/FastWeavTurb.

The vorticity field is generated in the periodic box. Its corresponding incompressible velocity field $\boldsymbol{u}=\mathcal{F}^{-1}\left({i \boldsymbol{k} \times \widehat{\boldsymbol{\omega}}}/{k^2}\right)$ is calculated by the spectral Biot-Savart law \citep{Shen2023Role,xiong2019Construction}, where $\mathcal{F}^{-1}$ denotes the operator of inverse Fourier transform and $\widehat{\boldsymbol{\omega}} = \mathcal{F}(\boldsymbol{\omega})$ the Fourier coefficient of $\boldsymbol{\omega}$ with the Fourier transform operator $\mathcal{F}$.

\begin{algorithm}[!ht]
\renewcommand{\algorithmicrequire}{\textbf{Input:}}
\renewcommand{\algorithmicensure}{\textbf{Output:}}
\caption{Construction of woven turbulence with multi-scale vortices}
\label{algorithm}
\begin{algorithmic}[1] 

\REQUIRE  
    $H$, $r_{n}$, $r_{L}$, $r_{\Gamma}$, $r_{\sigma}$, $\mathcal{N}$, $\sigma_{\mathcal{N}}$, $n_{1}$, $\Gamma_{1}$, $\lambda_{\sigma}$, $\varrho$ and $N$
    ; 
\ENSURE $\boldsymbol{\omega}$;

\FOR{${i} \gets 1$ to $\mathcal{N}$}
    \STATE Calculate $L_{i}$ by \eqref{eq:vortex density};
    \STATE Calculate $\mathscr{N}$ by \eqref{eq:centerline_length};
    \FOR{${j} \gets 1$ to $n_{i}$}
        \STATE Generate $\boldsymbol{B}({J})$;
        \FOR{${J} \gets 1$ to $\mathscr{N}$}
            \STATE Calculate ${\mathcal{S}}_{J}$ by \eqref{eq:norm_disc_para};
            \STATE Calculate $\boldsymbol{A}_{J}$,$\boldsymbol{B}_{J}$,$\boldsymbol{C}_{J}$,$\boldsymbol{D}_{J}$,$\boldsymbol{E}_{J}$ and $\boldsymbol{F}_{J}$ by \eqref{eq:spline parameters};
            \STATE Obtain $\boldsymbol{c}(\mathcal{S})$ by \eqref{eq:spline};
            \STATE Calculate $N_{C}$ by \eqref{eq:NC};
            \FOR{${K} \gets 1$ to $N_{C}$}
                \STATE Calculate $\boldsymbol{c}_{{J},{K}}$ by \eqref{eq:dividing points};
                \STATE Calculate $\boldsymbol{T}_{{J},{K}}$ by \eqref{eq:Frenet};
                \STATE Calculate $i_{x,\mathrm{min}}$,$i_{x,\mathrm{max}}$,$i_{y,\mathrm{min}}$,$i_{y,\mathrm{max}}$,$i_{z,\mathrm{min}}$,$i_{z,\mathrm{max}}$ by \eqref{eq:box};
                \FOR{$i_{x} \gets i_{x,\mathrm{min}}$ to $i_{x,\mathrm{max}}$}
                    \FOR{$i_{y} \gets i_{y,\mathrm{min}}$ to $i_{y,\mathrm{max}}$}
                        \FOR{$i_{z} \gets i_{z,\mathrm{min}}$ to $i_{z,\mathrm{max}}$}
                            \IF {$\boldsymbol{x}(i_{x},i_{y},i_{z}) \in {V}_{{J},{K}}$}
                                \STATE Calculate $\widetilde{\mathcal{S}}_{\boldsymbol{x}}$ by \eqref{eq:zeta_x};
                                \STATE Calculate $\widetilde{\boldsymbol{T}}_{\boldsymbol{x}}$, $\widetilde{\boldsymbol{N}}_{\boldsymbol{x}}$,
                                $\widetilde{\boldsymbol{B}}_{\boldsymbol{x}}$ by \eqref{eq:Frenet};
    
                                \STATE Calculate $\widetilde{\rho}_{\boldsymbol{x}}$ by \eqref{eq:rho_x};
                                \STATE Calculate $\widetilde{\boldsymbol{e}}_{s,\boldsymbol{x}}$, $\widetilde{\boldsymbol{e}}_{\rho,\boldsymbol{x}}$ and $\widetilde{\boldsymbol{e}}_{\theta,\boldsymbol{x}}$ by \eqref{eq:LCF_x};
                                \STATE Calculate $\boldsymbol{\omega}_{ij}(\boldsymbol{x})$ by \eqref{eq:w_x};
                            \ENDIF
                        \ENDFOR
                    \ENDFOR
                \ENDFOR
            \ENDFOR
        \ENDFOR
        \STATE $\boldsymbol{\omega} \leftarrow \boldsymbol{\omega}+\boldsymbol{\omega}_{ij}$
    \ENDFOR
\ENDFOR

\end{algorithmic}
\end{algorithm}

\section{Derivation and validation of the energy spectrum scaling model}
\label{app:spectrum}

\subsection{Derivation of the scaling law in the energy-containing range}
\label{app:energy-containing range}

We investigate the energy spectrum of a thin vortex tube to explain \eqref{eq:r_E H}. 
The thin vortex tube has circulation $\Gamma$, vanishing core size $R \to 0$, and a centerline $\mathcal{C}$ of length $L$. 
By \eqref{eq:w_single}, the vorticity of the thin vortex tube is $\boldsymbol{\omega}(\boldsymbol{x}(s,\rho,\theta))=\Gamma \boldsymbol{e}_{s} \delta(\rho)/(2\pi\rho)$, where $\delta(\rho)$ denotes the delta function that satisfies
$
\int_{0}^{\infty} {\delta(\rho)} \mathrm{d}\rho
= 1.
$
Then the vorticity in Fourier space is
\begin{equation}
\begin{split}
\hat{\boldsymbol{\omega}}(\boldsymbol{k})
=\frac{1}{2\pi}\int_{{V}} \boldsymbol{\omega}(\boldsymbol{x}) e^{-\mathrm{i}\boldsymbol{k}\cdot\boldsymbol{x}}\mathrm{d}\boldsymbol{x}
=\frac{\Gamma}{2\pi}\int_{\mathcal{C}}  \boldsymbol{e}_{s} e^{-\mathrm{i}\boldsymbol{k}\cdot\boldsymbol{c}(s)}\mathrm{d}s.
\end{split}
\end{equation}
Thus the enstrophy in Fourier space is
\begin{equation}
\label{eq:enstrophy_Fourier}
    \Omega(\boldsymbol{k}) = \frac{1}{2}\hat{\boldsymbol{\omega}}(\boldsymbol{k})\cdot \overline{\hat{\boldsymbol{\omega}}}(\boldsymbol{k})
    = \frac{\Gamma^2}{4\pi^2} \iint_{\mathcal{C}}  \boldsymbol{e}_s \cdot \boldsymbol{e}_{s^{\prime}} \exp({-\mathrm{i}\boldsymbol{k}\cdot(\boldsymbol{c}(s)-\boldsymbol{c}(s^{\prime}))})  \mathrm{d}s \mathrm{d}s^{\prime}
\end{equation}
with $s^{\prime} \neq s$. 

The centerline $\mathcal{C}$ is controlled by the discrete points of FBB.
For control point indices ${J}$ and ${J}^{\prime}$, the corresponding arc-lengths are $s\approx c_{H} \delta_B {J}$ and $s^{\prime}\approx c_{H} \delta_B {J}^{\prime}$, respectively, according to \eqref{eq:centerline_length}.
Considering the continuity of the centerline, the FBB scaling by \eqref{eq:FBB_scaling} leads to the scaling of $\mathcal{C}$ 
\begin{equation}
\label{eq:scaling_C_1}
    \langle (\boldsymbol{c}(s) - \boldsymbol{c}(s^{\prime}))^2 \rangle_{\mathcal{C}} 
    \approx \left\langle \left(\boldsymbol{B}\left(\left\lfloor\frac{s}{c_{H} \delta_B}\right\rceil\right) - \boldsymbol{B}\left(\left\lfloor\frac{s^{\prime}}{c_{H} \delta_B}\right\rceil\right)\right)^2 \right\rangle_{\mathscr{N}}
    \approx \delta_{\mathcal{C}}^2|s - s^{\prime}|^{2H}
\end{equation}
with  $\delta_{\mathcal{C}}=\delta_{B}/(\delta_{B}c_{H})^{H}$, 
where $\langle \cdot \rangle_{\mathcal{C}}$ denotes the average of fixed $|s - s^{\prime}|$ over $\mathcal{C}$.
Therefore, the correlation of tangent vectors is
\begin{equation}
\label{eq:scaling_C_2}
    \langle \boldsymbol{e}_s
    \cdot \boldsymbol{e}_{s^{\prime}} \rangle_{\mathcal{C}}
    =-\frac{1}{2}\frac{\partial^2\langle (\boldsymbol{c}(s) - \boldsymbol{c}(s^{\prime}))^2 \rangle_{\mathcal{C}}}{\partial s\partial s^{\prime}}
    \approx H(2H-1)\delta_{\mathcal{C}}^2|s - s^{\prime}|^{2H-2}.
\end{equation}
Then the Gaussian property of FBB \citep{Friedrich2020Stochastic} leads to 
\begin{equation}
\label{eq:FBB_Gaussian}
    \langle \exp(- \mathrm{i} \boldsymbol{k} \cdot (\boldsymbol{c}(s) - \boldsymbol{c}(s^{\prime}))) \rangle_{\mathcal{C}} \approx \exp\left(-\frac{k^2 \delta_{\mathcal{C}}^2|s - s^{\prime}|^{2H}}{6}\right).
\end{equation}

Supposing that $\boldsymbol{c}(s) - \boldsymbol{c}(s^{\prime})$ and $\boldsymbol{e}_s \cdot \boldsymbol{e}_{s^{\prime}}$ are uncorrelated, substituting \eqref{eq:scaling_C_1}, \eqref{eq:scaling_C_2}, and \eqref{eq:FBB_Gaussian} into \eqref{eq:enstrophy_Fourier} yields 
\begin{equation}
\begin{split}
    \Omega(\boldsymbol{k})
    &
    = \frac{\Gamma^2}{4\pi^2} \iint_{\mathcal{C}} \langle \boldsymbol{e}_s
    \cdot \boldsymbol{e}_{s^{\prime}} \rangle_{\mathcal{C}} \langle \exp(-i \boldsymbol{k} \cdot (\boldsymbol{c}(s) - \boldsymbol{c}(s^{\prime}))) \rangle_{\mathcal{C}}
    \mathrm{d}s \mathrm{d}s^{\prime}
    \\
    &
    \approx \frac{\Gamma^2}{4\pi^2} H(2H-1) \delta_{\mathcal{C}}^2 \iint_{\mathcal{C}} |s - s^{\prime}|^{2H-2}\exp\left(-\frac{k^2 \delta_{\mathcal{C}}^2|s - s^{\prime}|^{2H}}{6}\right)\mathrm{d}s \mathrm{d}s^{\prime}
    \\
    &
    = \frac{\Gamma^2}{4\pi^2} H(2H-1) k^{(1-2H)/H}\mathscr{E}
\end{split}
\end{equation}
with 
\begin{equation}
\label{eq:E_integral}
    \mathscr{E} = k^{-\frac{1}{H}}\int_{0}^{k^{1/H}\delta_{\mathcal{C}}^{{1}/{H}}L}\int_{0}^{k^{1/H}\delta_{\mathcal{C}}^{{1}/{H}}L} |t - t^{\prime}|^{2H-2} \exp\left(-\frac{|t - t^{\prime}|^{2H}}{6}\right) \mathrm{d}t \, \mathrm{d}t^{\prime},
\end{equation}
$t = k^{{1}/{H}} \delta_{\mathcal{C}}^{{1}/{H}} s$, and $t^{\prime} = k^{{1}/{H}} \delta_{\mathcal{C}}^{{1}/{H}} s^{\prime}$.
Since the integral in \eqref{eq:E_integral} is dominated by $|t - t^{\prime}| \ll 1$,
\begin{equation}
    \mathscr{E} \approx \frac{L}{H} \cdot 6^{\frac{2H-1}{2H}} \mathcal{G}\left(1 - \frac{1}{2H}\right)
\end{equation}
remains almost constant for positive integer $k$, where $\mathcal{G}(\cdot)$ denotes the Gamma function.
Thus the energy spectrum becomes 
\begin{equation}
\begin{split}
    E(k) 
    &
    = \frac{1}{k^2}\oint_{S(k)}\Omega(\boldsymbol{k})\mathrm{d}S(k)
    \propto k^{(1-2H)/H},
    \\
\end{split}
\end{equation}
i.e.~\eqref{eq:r_E H}.

\subsection{Validation of the inertial-range scaling model}
\label{app:inertial range}

\begin{figure}
  \centering
  \includegraphics[width=1.0\textwidth]{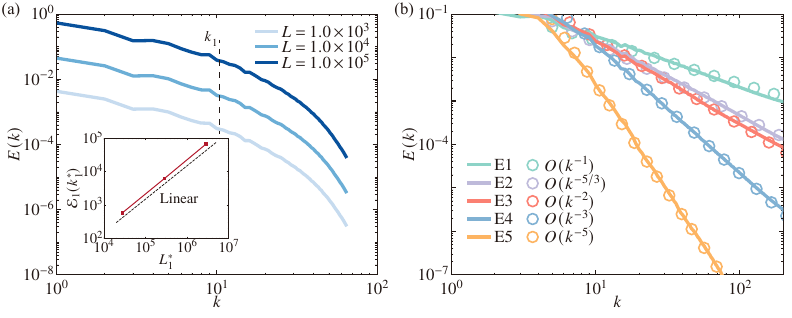}
  \caption{
  Numerical validations of (a) \eqref{eq:hyp_2} and (b) \eqref{eq:r_I}.
  (a) Energy spectra for cases in group D (see table~\ref{tab:set-up test cases}), where the dashed line denotes the location of the characteristic wavenumber $k_{1}$. The inset plots the variation of normalized characteristic energy $\mathcal{E}_{1}(k_{1}^{*})$ against normalized centerline length $L_{1}^{*}$.
  (b) Energy spectra for cases in group E (see table~\ref{tab:set-up test cases}).
  Solid lines represent numerical results and circles of corresponding colors denote theoretical predictions in \eqref{eq:r_I}.
  }
\label{fig:test}
\end{figure}

We constructed three vortex tubes of different lengths with $\mathcal{N}=1$ to verify the vortex-length-dependent energy hypothesis in \eqref{eq:hyp_2}. 
The specific settings of these cases in group D are provided in table~\ref{tab:set-up test cases}.
Changes in the centerline length do not alter the shape of the spectrum, but only alter the magnitude of the total energy, as shown in figure~\ref{fig:test}(a).
The normalized characteristic energy $\mathcal{E}_{1}(k_{1}^{*})$ and the normalized centerline length $L_{1}^{*}$ are linearly related, consistent with the prediction of \eqref{eq:hyp_2}. 
Additionally, the population $n_{i}$ and the centerline length $L_{i}$ play a similar role.
Thus, the normalized characteristic energy $\mathcal{E}_{i}(k_{i}^{*})$ is proportional to the total length $n_{i}L_{i}$ of vortex centerlines as suggested in \eqref{eq:hyp_2}.

\begin{table}
  \begin{center}
  \setlength{\tabcolsep}{6pt}
\def~{\hphantom{0}}
  \begin{tabular}{lccccccccccccc}
       Case & $r_{n}$ & $r_{L}$ & $r_{\Gamma}$ & $r_{\sigma}$ & $\mathcal{N}$ & $\sigma_{\mathcal{N}}$  & $\varrho$ & $N$ \\[3pt]
        D1   & - & - & - & - & 1 & $3.00\times 10^{-2}$ & $3.76\times 10^{-3}$ & 128 \\
        D2   & - & - & - & - & 1 & $3.00\times 10^{-2}$ & $3.76\times 10^{-2}$ & 128 \\
        D3   & - & - & - & - & 1 & $3.00\times 10^{-2}$ & $3.76\times 10^{-1}$ & 128 \\
        E1   & 8 & 1/4 & 0.707 & 1/2 & 8 & $1.7\times 10^{-3}$ & $3.97 \times 10^{1}$ & 512 \\
        E2   & 8 & 1/4 & 0.561 & 1/2 & 8 & $1.7\times 10^{-3}$ & $3.97 \times 10^{1}$ & 512 \\
        E3   & 8 & 1/2 & 1/4 & 1/4 & 4 & $1.7\times 10^{-3}$ & $3.31 \times 10^{0}$ & 512 \\
        E4   & 8 & 1/2 & 1/4 & 1/2 & 6 & $1.7\times 10^{-3}$ & $1.87 \times 10^{0}$ & 512 \\
        E5   & 8 & 1/8 & 1/4 & 1/2 & 8 & $1.7\times 10^{-3}$ & $2.66 \times 10^{1}$ & 512 \\
  \end{tabular}
  \caption{Set-up of supplementary woven turbulence cases. 
  Groups D and E are designed to validate \eqref{eq:hyp_2} and \eqref{eq:r_I}, respectively. 
  }
  \label{tab:set-up test cases}
  \end{center}
\end{table}

Furthermore, we tested the scaling exponent \eqref{eq:r_I} in inertial-range modelling under multiple self-similar ratios $r_{\sigma}$, $r_{\Gamma}$, and $r_{L}$, as shown in figure~\ref{fig:test}(b). 
Here we adjust the total number of sets $\mathcal{N}$ to ensure a sufficiently long inertial range, and the energy-containing and dissipation ranges are not shown for clarity. 
The prediction of the scaling exponent $r_{I}$ within the inertial range by \eqref{eq:r_I} agrees well with the numerical results.
Note that $r_{n}r_{L}$ acts as a combined parameter (not shown). 

Moreover, under special flow conditions, such as rotating turbulence \citep{Hu2022Transfer} and magnetohydrodynamics \citep{Yang2016Energy,Boldyrev2006Spectrum}, $r_{I}$ may deviate from -5/3. 
Therefore, the capability to precisely adjust $r_{I}$ over a wide range endows woven turbulence with a broad scope of applications.

\section{Randomness and spatial symmetry of woven turbulence}
\label{app:Rand_HIT}
We validate the randomness and spatial symmetry of woven turbulence to confirm its consistency with HIT. 
Spatial symmetry here refers to statistical translational and rotational invariance, corresponding to homogeneity and isotropy, respectively.
A priori analysis suggests that, although the vortex tubes do not fill the entire domain, a critical vortex density ensures that each tube and its surrounding zero-vorticity region acts as a representative unit.
The approximate random and uniform spatial distribution of these units implies that statistical randomness, homogeneity, and isotropy are expected to hold.

\begin{figure}
  \centering
  \includegraphics[width=1.0\textwidth]{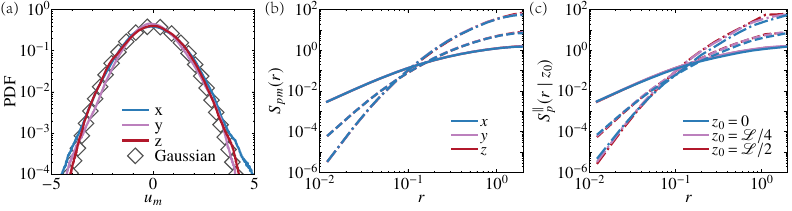}
  \caption{
  Assessment of randomness, homogeneity, and isotropy in woven turbulence, illustrated with case WT2-C.
  (a) PDFs of the velocity components
  $u_{x}=\boldsymbol{u}\cdot\boldsymbol{e}_{x}$,
  $u_{y}=\boldsymbol{u}\cdot\boldsymbol{e}_{y}$, 
  and $u_{z}=\boldsymbol{u}\cdot\boldsymbol{e}_{z}$,
  where $(\boldsymbol{e}_{x},\boldsymbol{e}_{y},\boldsymbol{e}_{z})$ denote unit vectors in the three coordinate directions $(x, y, z)$, respectively.
  (b) Directional structure functions $S_{px}(r) = \langle[(\boldsymbol{u}(\boldsymbol{x}+r\boldsymbol{e}_{x})-\boldsymbol{u}(\boldsymbol{x}))\cdot\boldsymbol{e}_{x}]^{p}\rangle$,
  $S_{py}(r) = \langle[(\boldsymbol{u}(\boldsymbol{x}+r\boldsymbol{e}_{y})-\boldsymbol{u}(\boldsymbol{x}))\cdot\boldsymbol{e}_{y}]^{p}\rangle$,
  and $S_{pz}(r) = \langle[(\boldsymbol{u}(\boldsymbol{x}+r\boldsymbol{e}_{z})-\boldsymbol{u}(\boldsymbol{x}))\cdot\boldsymbol{e}_{z}]^{p}\rangle$ in the three coordinate directions $(x, y, z)$.
  The solid line represents $p=2$, the dashed line $p=4$, and the dash-dotted line $p=6$.
  (c) Planar structure functions $S_{p}^{\parallel}(r \mid z_0) = \langle[(\boldsymbol{u}(\boldsymbol{x}+r\boldsymbol{e}_{m})-\boldsymbol{u}(\boldsymbol{x}))\cdot\boldsymbol{e}_{m}]^{p}\rangle_{z_{0}}$,
  where $\langle\cdot\rangle_{z = z_0}$ denotes the planar average over the two-dimensional slice at $z = z_0$.
  Line styles are the same as in (b).
  }
  \label{fig:Rand_space_sym}
\end{figure}

The fully random state of turbulence can be characterized by the Gaussian velocity distribution. 
By the Biot-Savart law, the velocity
\begin{equation}
    \label{eq:BS}
    \boldsymbol{u}(\boldsymbol{x})
    =\frac{1}{4\pi}\int^{{V}}  \frac{\boldsymbol{\omega}(\boldsymbol{x}^{\prime})\times(\boldsymbol{x}-\boldsymbol{x}^{\prime})}{|\boldsymbol{x}-\boldsymbol{x}^{\prime}|^3}\mathrm{d}V^{\prime}
    =\frac{1}{4\pi} \sum_{{i}=1}^{\mathcal{N}}
    \sum_{j=1}^{n_{i}}\sum_{J=1}^{\mathscr{N}} 
    \int^{{V}_{ijJ}}  \frac{\boldsymbol{\omega}_{ij}(\boldsymbol{x}^{\prime})\times(\boldsymbol{x}-\boldsymbol{x}^{\prime})}{|\boldsymbol{x}-\boldsymbol{x}^{\prime}|^3}\mathrm{d}V^{\prime}
\end{equation}
at position $\boldsymbol{x}$ is the sum of the induced velocities from multiple vortices, where
${V}_{ijJ}$ denotes the tubular region between $\boldsymbol{B}({J})$ and $\boldsymbol{B}({J}+1)$ of the ${j}$-th tube in set $i$, and $\mathrm{d}V^{\prime}$ denotes the volume element at position $\boldsymbol{x}^{\prime}$ in ${V}_{ijJ}$.
The FBB endows vortex tubes with sufficient randomness so that the induced velocities from different vortices in \eqref{eq:BS} can be approximated as independent and identically distributed (i.i.d.).
According to the central limit theorem \citep{E2019Applied, Durrett2010Probability}, the sum of i.i.d. random induced velocities tends toward the Gaussian distribution for sufficiently large vortex density $\varrho$. 
Therefore, as $\varrho$ increases, the PDF of velocity in \eqref{eq:BS} gradually converges to the Gaussian distribution in figure~\ref{fig:varrho_group}(b), which corresponds to numerous stochastic vortices in woven turbulence.
The critical vortex density in \eqref{eq:critical vortex density} balances the intermittency and Gaussian velocity distribution in woven turbulence.
Figure~\ref{fig:Rand_space_sym}(a) shows the PDFs of the velocity components in the $x$, $y$, and $z$ directions at this critical vortex density, all of which follow the Gaussian distribution, further verifying the randomness.

To assess the homogeneity and isotropy of woven turbulence, we examine the statistical distributions and structure functions across different directions and planes.
Figures \ref{fig:Rand_space_sym}(a) and (b) show the PDFs of the velocity components and directional structure functions in the three coordinate directions $(x, y, z)$, respectively, exhibiting good isotropy.
Figure~\ref{fig:Rand_space_sym}(c) shows the planar structure functions at different $z$-planes, indicating good spatial homogeneity in woven turbulence.
The small fluctuations at large scales in Figure~\ref{fig:Rand_space_sym}(b) and (c) indicate that the large-scale structures in woven turbulence are not perfectly homogeneous and isotropic, consistent with observations in real turbulence.

\section{DNS}
\label{app:DNS}

In the DNS, the unity density, incompressible NS equations
\begin{equation}
\label{eq:NS}
\begin{cases}\dfrac{\partial \boldsymbol{u}}{\partial t}+(\boldsymbol{u}\cdot\boldsymbol{\nabla})\boldsymbol{u}=-\boldsymbol{\nabla} p+\nu\nabla^2\boldsymbol{u}+\boldsymbol{f},
\\\boldsymbol{\nabla}\cdot\boldsymbol{u}=0\end{cases}
\end{equation}
are solved with different Reynolds numbers,
where $p$ denotes the pressure and $\boldsymbol{f}$ the external forcing for maintaining turbulence \citep{Machiels1997pre}. The forcing operates at large scales within a spherical shell of
radius $|\boldsymbol{k}|\leq 2$ in Fourier space.
Note that all quantities are dimensionless, and the characteristic length and velocity scales are $O(1)$.
The forced HIT is evolved from a Gaussian random velocity field with a specified energy spectrum $E(k,t=0) = c_E k^{-5/3}$. 
The constant $c_E$ is determined by the initial total energy $E_{t}= 3/2$ so that the root-mean-square velocity $u^{\prime}=1$.

The DNS was carried out using the standard pseudo-spectral method \citep{Rogallo1981Numerical, xiong2019Identifying}.
The computational domain is a periodic cube with side $\mathscr{L}=2\pi$ and discretized on $N^3$ uniform grid points.
The two-thirds truncation method with the maximum wavenumber $k_{\rm{max}} \approx N/3$ is used to remove aliasing errors. The second-order Adams-Bashforth scheme is employed for the time advancement. To ensure numerical stability and accuracy, the time step satisfies that the Courant-Friedrichs-Lewy number is less than $0.5$. 

We performed DNS of HIT at three different Reynolds numbers. 
The DNS data in the statistically stationary state are used in cases DNS1, DNS2, and DNS3 in table~\ref{tab:DNS cases}.   
The CPU time shown in figure~\ref{fig:case_group} corresponds to the temporal evolution up to $t=4$.
The DNS spatial resolution satisfies $k_{\rm{max}}\eta>1.5$ \citep{Pope2000Turbulent} for resolving the smallest scales.
Moreover, Case DNS4 at high $Re_\lambda$ is obtained from Johns Hopkins Turbulence Database \citep{Li2008A,Yeung2015Extreme}.

\bibliographystyle{jfm}
\bibliography{jfm}

\end{document}